\documentclass[12pt, a4paper]{article}
\usepackage[a4paper,margin=3cm]{geometry}
\usepackage{amsmath,amssymb}
\usepackage{authblk}
\usepackage{graphicx}
\usepackage{subcaption}
\usepackage{bm}
\usepackage{enumitem}
\usepackage{multirow}
\usepackage{tikz}
\usetikzlibrary{arrows.meta, positioning}
\usetikzlibrary{calc,intersections,through,backgrounds}
\usepackage[toc,page]{appendix}
\allowdisplaybreaks[2]
\usepackage{color}
\usepackage[multiple]{footmisc}
\usepackage{textcomp}

\usepackage{tikz-3dplot}
\usepackage{amscd}
\usepackage{amsthm}
\usepackage{array} 
\usepackage{graphicx}
\usepackage{adjustbox}
\usepackage{mathrsfs}
\usepackage{dsfont,comment} 
\usepackage[hidelinks]{hyperref}

\newcommand{\tx}[1]{\text {#1}}
\newcommand{\cl}[1]{\mathcal {#1}}

\newcommand{\bb}[1]{\mathbb {#1}}

\newcommand{\nd}{\noindent}

\newif\ifmn
\mntrue 
\definecolor{Ntxc}{rgb}{0.5,0.5,1}

\title{Finite group actions on genus two $SL(2,\mathbb C)$-character variety and applications to SCFTs}

\author{
	Semeon Arthamonov\thanks{E-mail: \href{mailto:arthamonov@bimsa.cn}{arthamonov@bimsa.cn}}
	\and 
	Anton Pribytok\thanks{E-mail: \href{mailto:antonspribitoks@bimsa.cn}{antonspribitoks@bimsa.cn}}
}

	\date{}

\begin{document}

\maketitle

\abstract{We investigate irreducible components of the fixed point sets of $ SL(2,\mathbb{C}) $-character variety of the genus two surface group under orientation preserving actions of the finite groups  of the $ Mod(\Sigma_{2}) $. We work in the \(\mathcal{O}\)-generator presentation of the genus two DAHA and its classical limit $ \mathcal{A}_{q=1,t} $, where we observe nontrivial coincidences between fixed loci attached to different subgroups and establish genus/irregularity transitions. The subvarieties obtained in this way provide novel geometric candidates for symmetry-reduced moduli spaces relevant to $ 4d $ $ \mathcal{N} = 2 $ SCFTs.} 

\newpage 
\tableofcontents

\section{Introduction}

\vspace{0.50cm}
Orientation preserving actions of finite subgroups on the surfaces of genus two and three were classified in \cite{Broughton'1991}. The presentation of the action in terms of Humphreys generators was subsequently constructed in \cite{NakamuraNakanishi'2018}.

On the other hand, in \cite{Arthamonov'2025} the first author, based on the previous work with Sh. Shakirov \cite{ArthamonovShakirov'2019}, has constructed a presentation in terms of generators and relations of an $SL(2,\mathbb C)$-character variety of the fundamental group of a genus two surface along with its flat Poisson deformation. Both the character variety and its deformation enjoy the action of the mapping class group of a genus two surface. 

This allows one to examine possibly reducible subvarieties of fixed points of the induced action on the $SL(2,\mathbb C)$-character variety and its deformation. In all but one cases we found decomposition of the varieties into irreducible components.

    \subsection{Character varieties.}
    
    Let $\Sigma_g$ be an oriented surface of genus $g$, its fundamental group has presentation with $2g$ generators and one relation
    \begin{equation}
    \pi_1(\Sigma_g)=\left\langle X_i,Y_i,\; 1\leq i\leq g\;\big|\; [X_1,Y_1]\dots[X_g,Y_g]=1\right\rangle.
    \label{eq:FundamentalGroupSigma_g}
    \end{equation}
    
    For the surface $\Sigma_g$ and complex affine algebraic group $G$ one can associate a \textit{representation variety} $\mathrm{Hom}(\pi_1(\Sigma_g),G)$ of the fundamental group $\pi_1(\Sigma_g)$. By definition, the coordinate ring of the representation variety of is obtained as a quotient of the $2g$ copies of the coordinate ring of the underlying algebraic group $G$ subject to additional relations inherited from the single relation (\ref{eq:FundamentalGroupSigma_g}) of the fundamental group
    \begin{equation}
    \mathbb C[\mathrm{Hom}(\pi_1(\Sigma_g),G)]:=\frac{\mathbb C[G]^{2g}}{(r_1,\dots, r_k)}.
    \label{eq:CoordinateRingRepresentationVariety}
    \end{equation}
    When $G=GL(n,\mathbb C)$ or $SL(n,\mathbb C)$, the only case we are interested in the current manuscript, ring (\ref{eq:CoordinateRingRepresentationVariety}) is known to be an integral domain (See, for example, Proposition 40 in \cite{Sikora'2012}). In this case, the coordinate ring has no zero divisors, and, in particular, no nilpotent elements.

    The group $G$ acts on the representation variety by simultaneous conjugation, this action is algebraic and gives rise to the action on the coordinate ring. Consider a $G$-invariant subring of the latter
    \begin{equation}
    \mathbb C[\mathrm{Hom}(\pi_1(\Sigma_g),G)]^G\;\subset\; \mathbb C[\mathrm{Hom}(\pi_1(\Sigma),G)],
    \label{eq:InvariantSubring}
    \end{equation}
    its spectrum is known as \textit{character variety} of the fundamental group $\pi_1(\Sigma_g)$.

    When $G$ is reductive, the invariant subring (\ref{eq:InvariantSubring}) must necessarily have presentation with finitely many generators and finitely many relations. However, finding such finite presentation explicitly is a notoriously hard problem from the computation standpoint for surfaces with $g>1$.

    \subsection{Mapping Class Group Action.}

    The quotient of an orientation preserving diffeomorphisms of $\Sigma_g$ by the diffeomorphisms isotopic to identity is known as the \textit{Mapping Class Group} of the surface
    \begin{equation*}
    \mathrm{Mod}(\Sigma_g):=\frac{Diff^{+}(\Sigma_g)}{Diff_0(\Sigma_g)}.
    \end{equation*}
    
\begin{figure}[h!]
    \centering
    \begin{tikzpicture}
        \node (left) {\includegraphics[height=3.5cm]{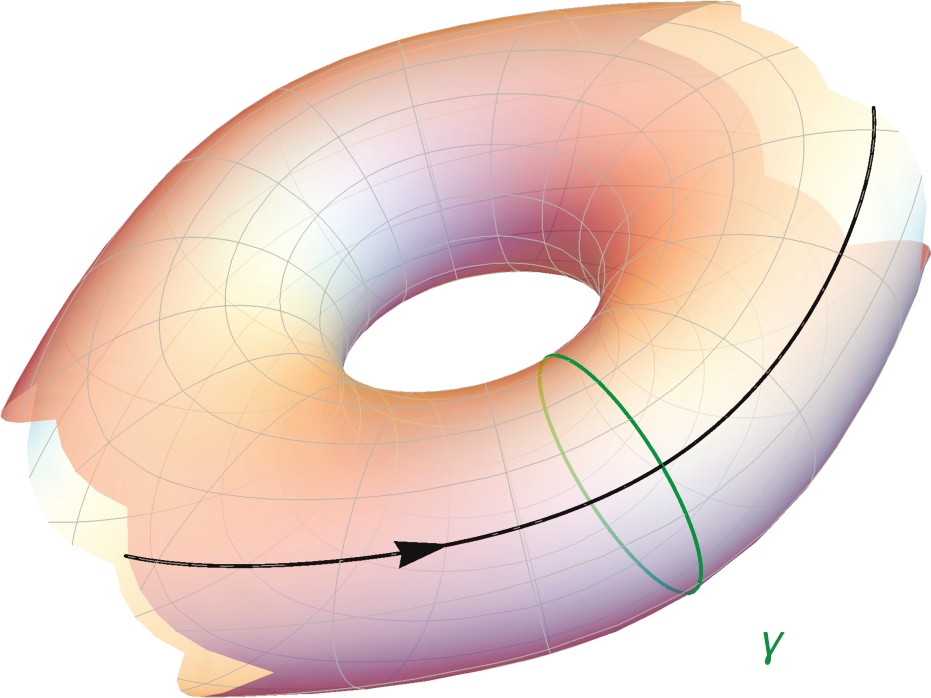}};
        \node (right) [right=1.5cm of left] {\includegraphics[height=3.5cm]{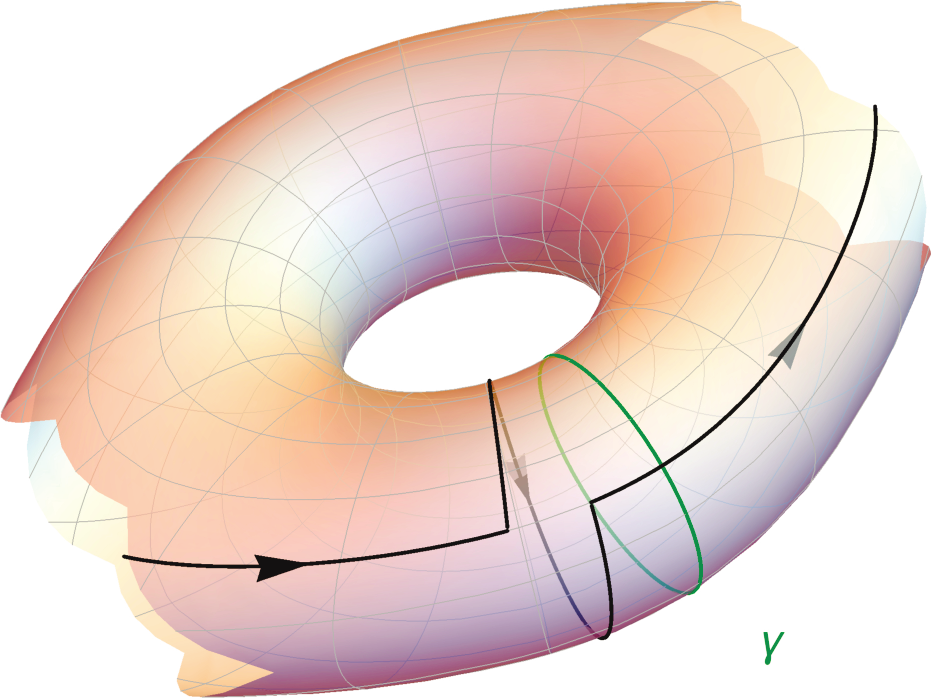}};
        \draw[{Bar[width=2mm]}-{Latex[length=2mm]}, thick]
        (left.east) -- (right.west)
        node[midway, above] {$D_\gamma$};
    \end{tikzpicture}
    \caption{Left Dehn twist along simple closed curve $\gamma$}
    \label{fig:LeftDehnTwist}
    \end{figure}

    The Mapping Class Group of a closed oriented surface is generated by finitely many Dehn twists \cite{Dehn'1938, Humphries'1979} (see Figure \ref{fig:LeftDehnTwist}). An explicit presentation with finitely many generators and finitely relations can be found in \cite{Wajnryb'1983}.

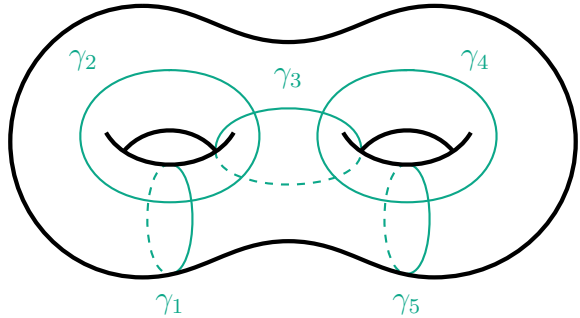
\begin{figure}[h!]
    \centering
    \begin{tikzpicture}[scale=1.2]
    \definecolor{DehnGreen}{RGB}{12,167,137}
    \draw[DehnGreen, thick, solid, looseness=0.7] (0.2,-0.25) to[out=0, in=0] (0.2,-1.39-0.07);
    \draw[DehnGreen, thick, dashed, looseness=0.7] (0.2,-0.25) to[out=180, in=180] (0.2,-1.39-0.07);
    \node[DehnGreen] at (0.2,-1.8) {$\gamma_1$};
    \draw[DehnGreen, thick, solid, looseness=2.3] (0.2,0.79) to[out=0,in=0] (0.2,-0.67) to[out=180,in=180] cycle;
    \node[DehnGreen] at (-0.75,0.9) {$\gamma_2$};
    \draw[DehnGreen, thick, solid] (2.8-0.5,-0.1) to[out=90, in=90] (0.2+0.5,-0.1);
    \draw[DehnGreen, thick, dashed] (0.2+0.5,-0.1) to[out=270,in=270,looseness=0.8] (2.8-0.5,-0.1);
    \node[DehnGreen] at (1.5,0.7) {$\gamma_3$};
    \draw[DehnGreen, thick, solid, looseness=2.3] (2.8,0.79) to[out=0,in=0] (2.8,-0.67) to[out=180,in=180] cycle;
    \node[DehnGreen] at (2.8+0.75,0.9) {$\gamma_4$};
    \draw[DehnGreen, thick, solid, looseness=0.7] (2.8,-0.25) to[out=0, in=0] (2.8,-1.39-0.07);
    \draw[DehnGreen, thick, dashed, looseness=0.7] (2.8,-0.25) to[out=180, in=180] (2.8,-1.39-0.07);
    \node[DehnGreen] at (2.8,-1.8) {$\gamma_5$};
    \draw[ultra thick] (-0.1,1.5) to[out=0,in=180] (1.5,1.1) to[out=0,in=180] (3.1,1.5) to[out=0,in=90] (4.65,0) to[out=270,in=0] (3.1,-1.5) to[out=180,in=0] (1.5,-1.1) to[out=180,in=0] (-0.1,-1.5) to[out=180,in=270] (-1.55,0) to[out=90,in=180] (-0.1,1.5);
    \draw[ultra thick] (0.2-0.5,-0.1) to[out=50,in=130] (0.2+0.5,-0.1);
    \draw[ultra thick] (0.2-0.7,0.1) to[out=-60,in=240] (0.2+0.7,0.1);
    \draw[ultra thick] (2.8-0.5,-0.1) to[out=50,in=130] (2.8+0.5,-0.1);
    \draw[ultra thick] (2.8-0.7,0.1) to[out=-60,in=240] (2.8+0.7,0.1);
    \end{tikzpicture}
    \caption{Simple closed curves corresponding to generators of $\mathrm{Mod}(\Sigma_2)$}
    \label{fig:GeneratorsGenusTwoMappingClassGroup}
    \end{figure}

    In particular, the mapping class group $\mathrm{Mod}(\Sigma_2)$ of the genus two surface is generated by five Dehn twists corresponding to simple closed curves depicted on Figure \ref{fig:GeneratorsGenusTwoMappingClassGroup}.

    Fix a choice of generic representative of simple closed curves $\gamma_1,\dots,\gamma_5\subset \Sigma_2$. Now let $p\in\Sigma_2\backslash(\gamma_1\cup\dots\cup\gamma_5)$ be a generic point outside of those arcs. Each of the left Dehn twists acts by automorphism of the fundamental group $\pi_1(\Sigma_2,p)$ based at $p$. This gives rise to the action of the mapping class group on $\mathbb C[\mathrm{Hom}(\pi_1(\Sigma_2,p),G)]^G$ by automorphisms of commutative ring.

    Now let $\Gamma\subset\mathrm{Mod}(\Sigma_2)$ be a finite subgroup of the mapping class group. We are interested in the subset of closed points of the character variety which are fixed by $\Gamma$. To this end we compute the vanishing ideal as a radical of the ideal generated by the constraints. In most cases, the resulting radical ideal would not be prime. We were able to compute the prime decomposition of all but one vanishing ideals and present the result in the following section.
    
    It is worth noting that the action of the Mapping Class Group on $SL(2,\mathbb C)$-character variety is not faithful, as the Hyperelliptic involution acts trivially. This naturally leads to the fact that $G_a$

    \section{\texorpdfstring{Classical Limit of DAHA and $ \bm{Mod(\Sigma_{2}} $) subgroups}{Classical Limit of DAHA and Mod(\Sigma_{2}) subgroups}}
	
	\vspace{0.50cm}
	\nd In this section we address the commutative classical limit \(q=1\) of the genus two DAHA and analyze the fixed loci of finite subgroups \(\Gamma\subset \mathrm{Mod}(\Sigma_2)\) in the \(\cl{O}\)-generator presentation. Since \(\cl{A}_{q=1,t}\) is a one parametric flat Poisson deformation of
    \begin{equation*}
		\cl{A}_{q=t=1}\cong \bb{C}\!\left[\mathrm{Hom}\!\bigl(\pi_{1}(\Sigma_{2}),SL(2,\bb{C})\bigr)\right]^{SL(2,\bb{C})} \,,
	\end{equation*} 
    where each subgroup action gives a system of algebraic constraints, whose radical ideal describes the corresponding classical and \(t\)-deformed fixed-point variety. 
	
	It turns out that for every $ Mod(\Sigma_{2}) $ subgroup as of \cite{Broughton'1991},  we can compute radical ideals, their primary decompositions and other characteristics both on the \(t=1\) fiber and in the \(t\)-deformed family. We also indicate the cases in which different subgroups produce equivalent varieties, either because their actions differ by the hyperelliptic involution $\zeta_{0}$ or appear to be mapped in rather nontrivial form. In the latter case, it involves transitions of genus and number of irregular points. In particular, the trivial \(G_a\) case reproduces the full \(6\)-dimensional variety, while the remaining cases give \(4\)- and \(2\)-dimensional families together with isolated zero-dimensional components.
	
	\subsection{\texorpdfstring{$\bm{G_{a}}$ subgroup}{G_a subgroup}}
    \(G_{a}\cong \mathbb{Z}_{2}\), \(|G_{a}|=2\), data \(\{0;\{2,6\}\}\) acts by the twist $\zeta_{0}$, which corresponds to an automorphism of the initial genus two DAHA and leaves all generators $\cl{O}$ invariant, hence reproduces the whole 6-dim DAHA variety.

	\subsection{\texorpdfstring{$\bm{G_{b}}$ or $\bm{G_{f}}$ subgroup}{G_{b} or G_{f} subgroup}}
	
	\(G_{b}\cong \mathbb{Z}_{2}\), \(|G_{b}|=2\), data \(\{1;\{2,2\}\}\):

    \label{sec:GbGf}

    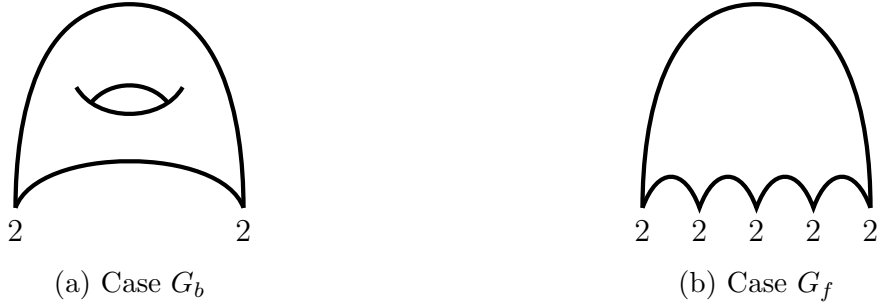
\begin{figure}[h!]
    \begin{subfigure}{0.45\linewidth}
    \centering
	\begin{tikzpicture}
    \draw[ultra thick] (-0.5,-0.1) to[out=50,in=130] (0.5,-0.1);
    \draw[ultra thick] (-0.7,0.1) to[out=-60,in=240] (0.7,0.1);
    \draw[ultra thick] (-1.5,-1.5) to[out=90,in=180] (0,1.2) to[out=0,in=90] (1.5,-1.5);
    \draw[ultra thick] (-1.5,-1.5) to[out=70,in=110, looseness=0.75] (1.5,-1.5);
    \node at (-1.5,-1.8) {$2$};
    \node at (1.5,-1.8) {$2$}; 
    \end{tikzpicture}
    \caption{Case $G_b$}
    \end{subfigure}
    \hfill
    \begin{subfigure}{0.45\linewidth}
     \centering
	\begin{tikzpicture}
    \draw[ultra thick] (-1.5,-1.5) to[out=90,in=180] (0,1.2) to[out=0,in=90] (1.5,-1.5);
    \draw[ultra thick, looseness=2] (-1.5,-1.5) to[out=70,in=110] (-0.75,-1.5) to[out=70,in=110] (0,-1.5) to[out=70,in=110] (0.75,-1.5) to[out=70,in=110] (1.5,-1.5);
    \node at (-1.5,-1.8) {$2$};
    \node at (-0.75,-1.8) {$2$};
    \node at (0,-1.8) {$2$};
    \node at (0.75,-1.8) {$2$};
    \node at (1.5,-1.8) {$2$}; 
    \end{tikzpicture}
    \caption{Case $G_f$}
    \end{subfigure}
    \caption{Branching data of the covering maps in Section \ref{sec:GbGf}.}
    \end{figure}

	\begin{equation}\label{eq:Gf}
		I_{b}:\qquad
		\begin{aligned}
			&\mathcal{O}_{4}=\mathcal{O}_{1},\quad
			\mathcal{O}_{5}=\mathcal{O}_{2},\quad
			\mathcal{O}_{6}=\mathcal{O}_{3},\quad
			\mathcal{O}_{4,5}=\mathcal{O}_{1,2},\quad
			\mathcal{O}_{5,6}=\mathcal{O}_{2,3},\quad
			\mathcal{O}_{6,1}=\mathcal{O}_{3,4},\\[8pt]
			&(\mathcal{O}_{1}\mathcal{O}_{3}-\mathcal{O}_{1,2}\mathcal{O}_{2,3})\,\mathcal{O}_{3,4}
			-\mathcal{O}_{2}\bigl(2\mathcal{O}_{1}\mathcal{O}_{3}-\mathcal{O}_{1}\mathcal{O}_{1,2}-\mathcal{O}_{3}\mathcal{O}_{2,3}\bigr)=0,\\[6pt]
			&\mathcal{O}_{2}\mathcal{O}_{1,2,3}-\bigl(2\mathcal{O}_{2}-\mathcal{O}_{1}\mathcal{O}_{3}
			+\mathcal{O}_{1,2}\mathcal{O}_{2,3}\bigr)=0,\\[4pt]
			&(\mathcal{O}_{1}\mathcal{O}_{3}-\mathcal{O}_{1,2}\mathcal{O}_{2,3})\,(\mathcal{O}_{3,4,5}-2)
			+\mathcal{O}_{2}\,(\mathcal{O}_{3}-\mathcal{O}_{1,2})^{2}=0, \\[4pt]
			&(\mathcal{O}_{1}\mathcal{O}_{3}-\mathcal{O}_{1,2}\mathcal{O}_{2,3})\,\mathcal{O}_{2,3,4}
			+\mathcal{O}_{1}^{2}\mathcal{O}_{2}-2\mathcal{O}_{1}\mathcal{O}_{3}
			-2\mathcal{O}_{1}\mathcal{O}_{2}\mathcal{O}_{2,3}
			+2\mathcal{O}_{1,2}\mathcal{O}_{2,3}
			+\mathcal{O}_{2}\mathcal{O}_{2,3}^{2}=0, \\[10pt]
			&(\mathcal{O}_{1}\mathcal{O}_{3}-\mathcal{O}_{1,2}\mathcal{O}_{2,3})
			\Bigl(
			2\mathcal{O}_{2}\bigl(2+\mathcal{O}_{1}\mathcal{O}_{2}\mathcal{O}_{3}\bigr)
			-\mathcal{O}_{2}\bigl(\mathcal{O}_{1}^{2}+2\mathcal{O}_{3}\mathcal{O}_{1,2}+\mathcal{O}_{2,3}^{2}\bigr)
			+\bigl(2\mathcal{O}_{2}-\mathcal{O}_{1}\mathcal{O}_{3}\\
			&
			+\mathcal{O}_{1,2}\mathcal{O}_{2,3}\bigr)\mathcal{O}_{2,3,4}
			\Bigr)+2\mathcal{O}_{2}\Bigl(
			\mathcal{O}_{2}^{2}\bigl(-2\mathcal{O}_{1}\mathcal{O}_{3}+\mathcal{O}_{1}\mathcal{O}_{1,2}
			+\mathcal{O}_{3}\mathcal{O}_{2,3}\bigr)
			-\mathcal{O}_{2}\,(\mathcal{O}_{3}-\mathcal{O}_{1,2})^{2}
			\Bigr)=0.
		\end{aligned}
	\end{equation}

	\paragraph{t-deformation.}

	\begin{equation}\label{eq:Gf-t-min}
		I^{(t)}_{b}:\qquad
		\begin{aligned}
			&\mathcal{O}_{4}=\mathcal{O}_{1},\quad
			\mathcal{O}_{5}=\mathcal{O}_{2},\quad
			\mathcal{O}_{6}=\mathcal{O}_{3},\quad
			\mathcal{O}_{4,5}=\mathcal{O}_{1,2},\quad
			\mathcal{O}_{5,6}=\mathcal{O}_{2,3},\quad
			\mathcal{O}_{6,1}=\mathcal{O}_{3,4},\\[8pt]
			&\sqrt t\,\mathcal{O}_{2}\mathcal{O}_{1,2,3}
			-\sqrt t\,\mathcal{O}_{1}\mathcal{O}_{3}
			+\sqrt t\,\mathcal{O}_{1,2}\mathcal{O}_{2,3}
			-\mathcal{O}_{2}(1+t)=0,\\
			&\sqrt t\,\mathcal{O}_{3}\mathcal{O}_{2,3,4}
			-\sqrt t\,\mathcal{O}_{1}\mathcal{O}_{2}
			+\sqrt t\,\mathcal{O}_{2,3}\mathcal{O}_{3,4}
			-\mathcal{O}_{3}(1+t)=0,\\
			&\sqrt t\,\mathcal{O}_{1}\mathcal{O}_{3,4,5}
			-\sqrt t\,\mathcal{O}_{2}\mathcal{O}_{3}
			+\sqrt t\,\mathcal{O}_{1,2}\mathcal{O}_{3,4}
			-\mathcal{O}_{1}(1+t)=0,\\[10pt]
			&\bigl(
			\mathcal{O}_{1}\mathcal{O}_{2}\mathcal{O}_{1,2}
			+\mathcal{O}_{2}\mathcal{O}_{3}\mathcal{O}_{2,3}
			+\mathcal{O}_{1}\mathcal{O}_{3}\mathcal{O}_{3,4}
			-\mathcal{O}_{1,2}\mathcal{O}_{2,3}\mathcal{O}_{3,4}
			\bigr)\sqrt t
			-\mathcal{O}_{1}\mathcal{O}_{2}\mathcal{O}_{3}(1+t)=0.
		\end{aligned}
	\end{equation}
	
	\noindent As indicated for the case of $G_{a}$ subgroup, it can be shown that 4-dim radical ideals appear equivalent for the two subgroups $G_{b}$ and \(G_{f}\cong \mathbb{Z}_{2}\times\mathbb{Z}_{2}\) (\(|G_{f}|=4\), \(\{0;\{2,5\}\}\)), which is due to the twist action difference by $\zeta_{0}$, which is the automorphism for genus two DAHA. 
	
	\subsection{\texorpdfstring{$\bm{G_{c}}$ subgroup}{G_{c} subgroup}}
	
	(\(G_{c}\cong \mathbb{Z}_{3}\), \(|G_{c}|=3\), data \(\{0;\{3,4\}\}\)): \\

    \begin{figure}[h!]
     \centering
	\begin{tikzpicture}
    \draw[ultra thick] (-1.5,-1.5) to[out=90,in=180] (0,1.2) to[out=0,in=90] (1.5,-1.5);
    \draw[ultra thick, looseness=2] (-1.5,-1.5) to[out=70,in=110] (-0.5,-1.5) to[out=70,in=110] (0.5,-1.5) to[out=70,in=110] (1.5,-1.5);
    \node at (-1.5,-1.8) {$3$};
    \node at (-0.5,-1.8) {$3$};
    \node at (0.5,-1.8) {$3$};
    \node at (1.5,-1.8) {$3$}; 
    \end{tikzpicture}
    \caption{Branching data of the covering maps in case $G_c$}
    \end{figure}
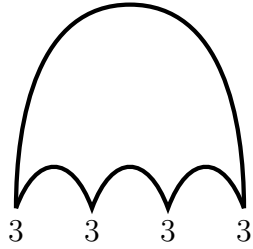
	
	
	\paragraph{$\bm{t=1}$:} $I_{1}  \, \cap \, I_{2}  \, \cap \, I_{3}$ with $\dim(I_{1}, I_{2}, I_{3}) = (2,0,0)$
	\begin{equation}
		\label{eq:I1-compact}
		I_{1}:\qquad
		\begin{aligned}
			&\mathcal{O}_{2}=\mathcal{O}_{4}=\mathcal{O}_{6},\qquad
			\mathcal{O}_{1,2}=\mathcal{O}_{3,4}=\mathcal{O}_{5,6},\\[2pt]
			&\mathcal{O}_{2,3}=\mathcal{O}_{4,5}=\mathcal{O}_{6,1},\qquad
			\mathcal{O}_{1,2,3}=\mathcal{O}_{2,3,4}=\mathcal{O}_{3,4,5},\\[6pt]
			&\mathcal{O}_{1}=\mathcal{O}_{3}=\mathcal{O}_{5}
			=-\frac{1}{3}\Bigl(
			3\mathcal{O}_{6} + \mathcal{O}_{3,4,5}^{3} +2\mathcal{O}_{3,4,5}^{2} \\ 
			& -\mathcal{O}_{3,4,5}\bigl(
			\mathcal{O}_{5,6}\mathcal{O}_{6,1}
			+\mathcal{O}_{5,6}
			+\mathcal{O}_{6,1}
			+3\bigr) - 6
			\Bigr) \\
			&\mathcal{O}_{6}^{2}-\mathcal{O}_{6}(\mathcal{O}_{3,4,5}+2)
			+\mathcal{O}_{5,6}+\mathcal{O}_{6,1}=0,\\[4pt]
			&\mathcal{O}_{5,6}(\mathcal{O}_{6,1}+1)+\mathcal{O}_{6,1}
			-\mathcal{O}_{3,4,5}(\mathcal{O}_{3,4,5}+2)=0 \,. \\[6pt]
		\end{aligned}
	\end{equation}
	\begin{equation}
		\begin{cases}
			I_{2}=\Bigl\langle\,
			\mathcal{O}_{i}+1, \; \mathcal{O}_{i \, i+1}+1, \, \mathcal{O}_{abc}-2
			\,\Bigr\rangle  \\
			I_{3}=\Bigl\langle\,
			\mathcal{O}_{i}, \, 
			\mathcal{O}_{i \, i+1} - 2, \,
			\mathcal{O}_{j,j+1,j+2}+1 \, \Bigr\rangle 
		\end{cases} \hspace{-5.00mm} , \quad
		\forall \, i = \overline{1, \, 6} \tx{ and } j = (1,2,3)_{\tx{cyc}}
	\end{equation}

	\paragraph{t-deformed:} $I_{1}  \, \cap \, I_{2} $ with $\dim(I_{1}, I_{2}) = (2,0)$ 
	\begin{equation}\label{eq:Gc-t-I1}
		I^{(t)}_{1}:\qquad
		\begin{aligned}
			&\mathcal{O}_{1}=\mathcal{O}_{3}=\mathcal{O}_{5},\quad
			\mathcal{O}_{2}=\mathcal{O}_{4}=\mathcal{O}_{6}, \quad \mathcal{O}_{1,2}=\mathcal{O}_{3,4}=\mathcal{O}_{5,6}, \\
			&\mathcal{O}_{2,3}=\mathcal{O}_{4,5}=\mathcal{O}_{6,1}
			=\mathcal{O}_{1}\mathcal{O}_{2}-\mathcal{O}_{1,2}, \quad \mathcal{O}_{1,2,3}=\mathcal{O}_{2,3,4}=\mathcal{O}_{3,4,5}
			=\mathcal{O}_{1}+\mathcal{O}_{2}-\frac{1+t}{\sqrt{t}},\\[8pt]
			&\mathcal{O}_{1}+\mathcal{O}_{2}
			-\sqrt{t}\Bigl(
			\mathcal{O}_{1}^{2}+\mathcal{O}_{2}^{2}+\mathcal{O}_{1,2}^{2}
			+\mathcal{O}_{1}\mathcal{O}_{2}(1-\mathcal{O}_{1,2})
			\Bigr)
			+(\mathcal{O}_{1}+\mathcal{O}_{2})t=0.
		\end{aligned}
	\end{equation}

	\begin{equation}\label{eq:Gc-t-I2}
		I^{(t)}_{2}:\qquad
		\begin{aligned}
			&\mathcal{O}_{1}=\mathcal{O}_{2}=\mathcal{O}_{3}=\mathcal{O}_{4}=\mathcal{O}_{5}=\mathcal{O}_{6}
			=t^{-1/6}+t^{1/6},\\[4pt]
			&\mathcal{O}_{1,2}=\mathcal{O}_{2,3}=\mathcal{O}_{3,4}=\mathcal{O}_{4,5}
			=\mathcal{O}_{5,6}=\mathcal{O}_{6,1}\\
			&\hspace{0.75cm}
			=\frac12\Bigl(2+(1+i\sqrt3)\,t^{-1/3}+(1-i\sqrt3)\,t^{1/3}\Bigr),\\[6pt]
			&\mathcal{O}_{1,2,3}=\mathcal{O}_{2,3,4}=\mathcal{O}_{3,4,5}
			=\frac{(1+t^{1/3})\bigl(1+i\sqrt3+(1-i\sqrt3)t^{2/3}\bigr)}{2\,t^{1/2}}.
		\end{aligned}
	\end{equation}

	\subsection{\texorpdfstring{$\bm{G_{e}}$ subgroup}{G_{e} subgroup}}
	
	\(G_{e}\cong \mathbb{Z}_{4}\), \(|G_{e}|=4\), data \(\{0;\{2,2\},\{4,2\}\}\). \\

     \begin{figure}[h!]
     \centering
	\begin{tikzpicture}
    \draw[ultra thick] (-1.5,-1.5) to[out=90,in=180] (0,1.2) to[out=0,in=90] (1.5,-1.5);
    \draw[ultra thick, looseness=2] (-1.5,-1.5) to[out=70,in=110] (-0.5,-1.5) to[out=70,in=110] (0.5,-1.5) to[out=70,in=110] (1.5,-1.5);
    \node at (-1.5,-1.8) {$2$};
    \node at (-0.5,-1.8) {$2$};
    \node at (0.5,-1.8) {$4$};
    \node at (1.5,-1.8) {$4$}; 
    \end{tikzpicture}
    \caption{Branching data of the covering maps in case $G_e$}
    \end{figure}
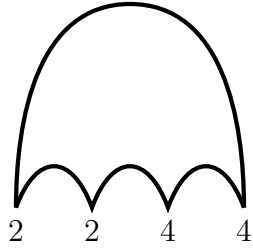
	
	$I_{1}  \, \cap \, I_{2}  $ with $\dim(I_{1}, I_{2}) = (2,2)$: 
	\begin{equation}\label{eq:I1-Ge-compact}
		I_{1} :\qquad
		\begin{aligned}
			&\mathcal{O}_{4}=\mathcal{O}_{2},\qquad
			\mathcal{O}_{5}=\mathcal{O}_{1},\qquad
			\mathcal{O}_{6,1}=\mathcal{O}_{4,5},\qquad
			\mathcal{O}_{3,4,5}=\mathcal{O}_{1},\\[4pt]
			&\mathcal{O}_{6}=2(\mathcal{O}_{2}-1), \quad \mathcal{O}_{1,2}=\mathcal{O}_{1}\mathcal{O}_{2}-\mathcal{O}_{4,5},\qquad
			\mathcal{O}_{5,6}=2(\mathcal{O}_{2}-1)\mathcal{O}_{1}-\mathcal{O}_{4,5},\\
			&\mathcal{O}_{3,4}=\frac{\mathcal{O}_{1}^{2}}{2},\qquad
			\mathcal{O}_{2,3}=\frac{\mathcal{O}_{4,5}^{2}}{2},\\
			&\mathcal{O}_{1,2,3}=\mathcal{O}_{1}+(\mathcal{O}_{2}-2)\mathcal{O}_{4,5},\qquad
			\mathcal{O}_{2,3,4}=\mathcal{O}_{3}+2(\mathcal{O}_{2}-2),\\[6pt]
			&2\mathcal{O}_{2}\mathcal{O}_{3}-\mathcal{O}_{1}^{2}-\mathcal{O}_{4,5}^{2}=0, \quad \mathcal{O}_{1}^{2}+\mathcal{O}_{4,5}^{2}+2\mathcal{O}_{2}^{2}
			-\mathcal{O}_{1}\mathcal{O}_{2}\mathcal{O}_{4,5}-4\mathcal{O}_{2}=0.
		\end{aligned}
	\end{equation}

	\begin{equation}\label{eq:I2-Ge-compact}
		I_{2}:\qquad
		\begin{aligned}
			&\mathcal{O}_{4}=\mathcal{O}_{2},\qquad
			\mathcal{O}_{5}=-\mathcal{O}_{1},\qquad
			\mathcal{O}_{6,1}=\mathcal{O}_{4,5},\qquad
			\mathcal{O}_{3,4,5}=-\mathcal{O}_{1},\\[4pt]
			&\mathcal{O}_{6}=-2(1+\mathcal{O}_{2}), \quad \mathcal{O}_{1,2}=\mathcal{O}_{1}\mathcal{O}_{2}+\mathcal{O}_{4,5},\qquad
			\mathcal{O}_{5,6}=\mathcal{O}_{4,5}+2\mathcal{O}_{1}(1+\mathcal{O}_{2}),\\
			&2\mathcal{O}_{3,4}-\mathcal{O}_{1}^{2}=0,\qquad
			2\mathcal{O}_{2,3}-\mathcal{O}_{4,5}^{2}=0,\\[4pt]
			&2\mathcal{O}_{2}\mathcal{O}_{3}-\mathcal{O}_{1}^{2}-\mathcal{O}_{4,5}^{2}=0, \quad 2\mathcal{O}_{2}\mathcal{O}_{2,3,4}-\bigl(\mathcal{O}_{1}^{2}+8\mathcal{O}_{2}+4\mathcal{O}_{2}^{2}
			+\mathcal{O}_{4,5}^{2}\bigr)=0,\\
			&2\mathcal{O}_{2}\mathcal{O}_{1,2,3}-\bigl(2\mathcal{O}_{1}\mathcal{O}_{2}
			+\mathcal{O}_{1}^{2}\mathcal{O}_{4,5}
			+\mathcal{O}_{1}\mathcal{O}_{2}\mathcal{O}_{4,5}^{2}
			+\mathcal{O}_{4,5}^{3}\bigr)=0,\\
			&\mathcal{O}_{1}^{2}+\mathcal{O}_{4,5}^{2}+4\mathcal{O}_{2}+2\mathcal{O}_{2}^{2}
			+\mathcal{O}_{1}\mathcal{O}_{2}\mathcal{O}_{4,5}=0.
		\end{aligned}
	\end{equation}

	\paragraph{t-deformed}
	
	\begin{equation}\label{eq:Ge-t-I1}
		I^{(t)}_{1}:\qquad
		\begin{aligned}
			&\mathcal{O}_{4}=\mathcal{O}_{2},\qquad
			\mathcal{O}_{5}=\mathcal{O}_{1},\qquad
			\mathcal{O}_{6,1}=\mathcal{O}_{4,5},\qquad
			\mathcal{O}_{3,4,5}=\mathcal{O}_{1},\\[6pt]
			&\sqrt t\,\mathcal{O}_{6}-2\mathcal{O}_{2}\sqrt t+(1+t)=0, \quad \mathcal{O}_{1,2}-\bigl(\mathcal{O}_{1}\mathcal{O}_{2}-\mathcal{O}_{4,5}\bigr)=0,\\
			&2\mathcal{O}_{2,3}-\mathcal{O}_{4,5}^{2}=0,\qquad
			2\mathcal{O}_{3,4}-\mathcal{O}_{1}^{2}=0,\qquad
			2\mathcal{O}_{2}\mathcal{O}_{3}-\mathcal{O}_{1}^{2}-\mathcal{O}_{4,5}^{2}=0,\\[6pt]
			&\sqrt t\,\bigl(\mathcal{O}_{5,6}+\mathcal{O}_{4,5}-2\mathcal{O}_{1}\mathcal{O}_{2}\bigr)
			+\mathcal{O}_{1}(1+t)=0,\\
			&2\mathcal{O}_{2}\mathcal{O}_{1,2,3}
			+\mathcal{O}_{1}^{2}\mathcal{O}_{4,5}+\mathcal{O}_{4,5}^{3}
			-\mathcal{O}_{1}\mathcal{O}_{2}\bigl(2+\mathcal{O}_{4,5}^{2}\bigr)=0,\\
			&2\mathcal{O}_{2}\sqrt t\,\mathcal{O}_{2,3,4}
			-\bigl(\mathcal{O}_{1}^{2}+4\mathcal{O}_{2}^{2}+\mathcal{O}_{4,5}^{2}\bigr)\sqrt t
			+4\mathcal{O}_{2}(1+t)=0,\\[10pt]
			&2\mathcal{O}_{2}(1+t)
			-\sqrt t\,\bigl(\mathcal{O}_{1}^{2}+2\mathcal{O}_{2}^{2}
			-\mathcal{O}_{1}\mathcal{O}_{2}\mathcal{O}_{4,5}+\mathcal{O}_{4,5}^{2}\bigr)=0.
		\end{aligned}
	\end{equation}

	\begin{equation}\label{eq:Ge-t-I2-corrected}
		I^{(t)}_{2}:\qquad
		\begin{aligned}
			&\mathcal{O}_{4}=\mathcal{O}_{2},\qquad
			\mathcal{O}_{5}=-\mathcal{O}_{1},\qquad
			\mathcal{O}_{6,1}=\mathcal{O}_{4,5},\qquad
			\mathcal{O}_{3,4,5}=-\mathcal{O}_{1},\\[6pt]
			&\sqrt t\,\mathcal{O}_{6}+\bigl(1+2\mathcal{O}_{2}\sqrt t+t\bigr)=0, \quad \mathcal{O}_{1,2}-\bigl(\mathcal{O}_{1}\mathcal{O}_{2}+\mathcal{O}_{4,5}\bigr)=0,\\
			&2\mathcal{O}_{2,3}-\mathcal{O}_{4,5}^{2}=0,\qquad
			2\mathcal{O}_{3,4}-\mathcal{O}_{1}^{2}=0,\qquad
			2\mathcal{O}_{2}\mathcal{O}_{3}-\mathcal{O}_{1}^{2}-\mathcal{O}_{4,5}^{2}=0,\\[6pt]
			&\sqrt t\,(\mathcal{O}_{5,6}-\mathcal{O}_{4,5})
			-\mathcal{O}_{1}\bigl(1+2\mathcal{O}_{2}\sqrt t+t\bigr)=0,\\
			&2\mathcal{O}_{2}\mathcal{O}_{1,2,3}
			-\Bigl(\mathcal{O}_{1}^{2}\mathcal{O}_{4,5}+\mathcal{O}_{4,5}^{3}
			+\mathcal{O}_{1}\mathcal{O}_{2}(2+\mathcal{O}_{4,5}^{2})\Bigr)=0,\\
			&2\mathcal{O}_{2}\sqrt t\,\mathcal{O}_{2,3,4}
			-\bigl(\mathcal{O}_{1}^{2}+4\mathcal{O}_{2}^{2}+\mathcal{O}_{4,5}^{2}\bigr)\sqrt t
			-4\mathcal{O}_{2}(1+t)=0,\\[10pt]
			&\sqrt t\,\bigl(\mathcal{O}_{1}^{2}+2\mathcal{O}_{2}^{2}
			+\mathcal{O}_{1}\mathcal{O}_{2}\mathcal{O}_{4,5}+\mathcal{O}_{4,5}^{2}\bigr)
			+2\mathcal{O}_{2}(1+t)=0.
		\end{aligned}
	\end{equation}

	\begin{equation}\label{eq:Ge-t-I3}
		I^{(t)}_{3}:\qquad
		\begin{aligned}
			&\mathcal{O}_{2}=\mathcal{O}_{3}=\mathcal{O}_{4}=2, \quad \mathcal{O}_{6}=\frac{1+t}{\sqrt{t}}, \quad \mathcal{O}_{2,3}=\mathcal{O}_{3,4}
			=2-\frac{i(t-1)}{\sqrt{t}},\\[6pt]
			&\mathcal{O}_{1}=\mathcal{O}_{1,2}
			=-\frac{(1+i)(\sqrt{t}-i)}{t^{1/4}}, \quad \mathcal{O}_{5,6}
			=-\frac{(1-i)(i+t^{3/2})}{t^{3/4}} \\
			&\mathcal{O}_{5}=\mathcal{O}_{4,5}=\mathcal{O}_{6,1}
			=-\frac{(1-i)(\sqrt{t}+i)}{t^{1/4}} \,. \\[6pt]
		\end{aligned}
	\end{equation}

	\begin{equation}\label{eq:Ge-t-I4}
		I^{(t)}_{4}:\qquad
		\begin{aligned}
			&\mathcal{O}_{2}=\mathcal{O}_{3}=\mathcal{O}_{4}=-2, \quad \mathcal{O}_{6}=\frac{1+t}{\sqrt{t}}, \quad \mathcal{O}_{2,3}=\mathcal{O}_{3,4}
			=2-\frac{i(t-1)}{\sqrt{t}},\\[6pt]
			&\mathcal{O}_{1}=-\mathcal{O}_{1,2}
			=\frac{(1+i)(\sqrt{t}-i)}{t^{1/4}}, \quad \mathcal{O}_{5,6}
			=-\frac{(1-i)(i+t^{3/2})}{t^{3/4}}, \\[6pt]
			&-\mathcal{O}_{5}=\mathcal{O}_{4,5}=\mathcal{O}_{6,1}
			=\frac{(1-i)(\sqrt{t}+i)}{t^{1/4}} \,. \\[6pt]
		\end{aligned}
	\end{equation}

	\subsection{\texorpdfstring{$\bm{G_{h}}$ or $\bm{G_{o}}$ subgroup}{G_{h} or G_{o} subgroup}}
    \(G_{h}\cong \mathbb{Z}_{5}\), \(|G_{h}|=5\), data \(\{0;\{5,3\}\}\) 

    \label{sec:GhGo}

     \begin{figure}[h!]
     \begin{subfigure}{0.45\linewidth}
     \centering
	\begin{tikzpicture}
    \draw[ultra thick] (-1.5,-1.5) to[out=90,in=180] (0,1.2) to[out=0,in=90] (1.5,-1.5);
    \draw[ultra thick, looseness=2] (-1.5,-1.5) to[out=70,in=110] (0,-1.5) to[out=70,in=110] (1.5,-1.5);
    \node at (-1.5,-1.8) {$5$};
    \node at (-0,-1.8) {$5$};
    \node at (1.5,-1.8) {$5$}; 
    \end{tikzpicture}
    \caption{Case $G_h$.}
    \end{subfigure}
    \hfill
    \begin{subfigure}{0.45\linewidth}
     \centering
	\begin{tikzpicture}
    \draw[ultra thick] (-1.5,-1.5) to[out=90,in=180] (0,1.2) to[out=0,in=90] (1.5,-1.5);
    \draw[ultra thick, looseness=2] (-1.5,-1.5) to[out=70,in=110] (0,-1.5) to[out=70,in=110] (1.5,-1.5);
    \node at (-1.5,-1.8) {$2$};
    \node at (-0,-1.8) {$5$};
    \node at (1.5,-1.8) {$10$}; 
    \end{tikzpicture}
    \caption{Case $G_o$.}
    \end{subfigure}
    \caption{Branching data of the covering maps in case Section \ref{sec:GhGo}.}
    \end{figure}
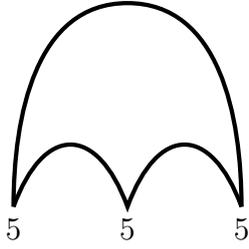
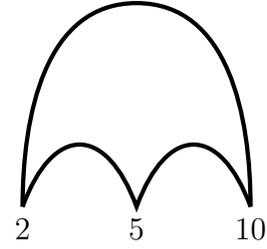
	
	\paragraph{$\bm{t=1}$:} $I_{1} \, \cap \, I_{2} \, \cap \, I_{3}$ with $\dim(I_{1}, I_{2}, I_{3}) = (0,0,0)$

	\begin{equation}
		I_{1}=\Bigl\langle\, \mathcal{O}-2 \,, \ \forall\,\mathcal{O}\,\Bigr\rangle
		\qquad 
		I_{2}:\quad
		\begin{aligned}
			&\mathcal{O}_{1}=\mathcal{O}_{2}=\mathcal{O}_{3}=\mathcal{O}_{4}
			=\mathcal{O}_{1,2,3}=\mathcal{O}_{2,3,4}
			=\frac{-1+\sqrt{5}}{2},\\
			&\mathcal{O}_{5}=\mathcal{O}_{6}=\mathcal{O}_{4,5}
			=\mathcal{O}_{6,1}=\mathcal{O}_{3,4,5}
			=\frac{-1-\sqrt{5}}{2},\\
			&\mathcal{O}_{1,2}=\mathcal{O}_{2,3}=\mathcal{O}_{3,4}
			=\mathcal{O}_{5,6}=2.
		\end{aligned}
	\end{equation} 
	\begin{equation}\label{eq:Gh-I3}
		I_{3}:\qquad
		\begin{aligned}
			&\mathcal{O}_{1}=\mathcal{O}_{2}=\mathcal{O}_{3}=\mathcal{O}_{4}
			=2+\sqrt{5},\\
			&\mathcal{O}_{5}=\mathcal{O}_{6}=\mathcal{O}_{4,5}
			=\mathcal{O}_{6,1}=\mathcal{O}_{3,4,5}
			=-3-\sqrt{5},\\
			&\mathcal{O}_{1,2}=\mathcal{O}_{2,3}=\mathcal{O}_{3,4}
			=\frac{9+5\sqrt{5}}{2},\qquad
			\mathcal{O}_{5,6}=12+5\sqrt{5},\\
			&\mathcal{O}_{1,2,3}=\mathcal{O}_{2,3,4}
			=\frac{19+7\sqrt{5}}{2}.
		\end{aligned}
	\end{equation}
	
	\nd The equivalent variety structure arises for the case of $G_{o}$ subgroup, which is given by (\(G_{o}\cong \mathbb{Z}_{10}\), \(|G_{o}|=10\), data \(\{0;\{2,1\},\{5,1\},\{10,1\}\}\)).
	
	\paragraph{t-deformed:} $I_{1}$ with $\dim(I_{1}) = 0$

	\begin{equation}\label{eq:Gh-t-I1}
		I_{1}^{(t)}:\quad
		\begin{aligned}
			&\mathcal{O}_{2}=\mathcal{O}_{3}=\mathcal{O}_{4}=\mathcal{O}_{1}, \quad \mathcal{O}_{6}=\mathcal{O}_{4,5}=\mathcal{O}_{6,1}=\mathcal{O}_{3,4,5}
			=\mathcal{O}_{5} \\
			&2\mathcal{O}_{1}\mathcal{O}_{1,2}
			=2\mathcal{O}_{1}\mathcal{O}_{2,3}
			=2\mathcal{O}_{1}\mathcal{O}_{3,4}
			=(\mathcal{O}_{1}-1)\mathcal{O}_{1}^{2}+\mathcal{O}_{5}^{2},\\
			&\mathcal{O}_{5,6}=-\mathcal{O}_{1}+\mathcal{O}_{5}^{2},\\
			&2\mathcal{O}_{1}\mathcal{O}_{1,2,3}
			=2\mathcal{O}_{1}\mathcal{O}_{2,3,4}
			=\mathcal{O}_{1}^{2}-\mathcal{O}_{1}^{3}
			+2\mathcal{O}_{1}\mathcal{O}_{5}^{2}-\mathcal{O}_{5}^{2},\\[8pt]
			&2\mathcal{O}_{1}^{2}
			+\mathcal{O}_{5}\bigl((\mathcal{O}_{1}-5)\mathcal{O}_{1}^{2}+\mathcal{O}_{5}^{2}\bigr)t^{1/4}
			+2\mathcal{O}_{1}^{2}\sqrt{t}=0,\\
			&(\,3\mathcal{O}_{1}-5\,)\mathcal{O}_{5}^{2}t^{1/4}
			=2\mathcal{O}_{5}(1+\sqrt{t})
			+\mathcal{O}_{1}^{2}(\mathcal{O}_{1}^{2}-5)t^{1/4},\\
			&4\mathcal{O}_{1}^{2}\mathcal{O}_{5}
			+\bigl(-5\mathcal{O}_{1}^{4}+\mathcal{O}_{1}^{6}
			-2\mathcal{O}_{1}^{3}\mathcal{O}_{5}^{2}+\mathcal{O}_{5}^{4}\bigr)t^{1/4}
			+4\mathcal{O}_{1}^{2}\mathcal{O}_{5}\sqrt{t}=0,\\
			&\mathcal{O}_{1}\mathcal{O}_{5}t^{1/4}
			+\bigl(-2+(5+(\mathcal{O}_{1}-5)\mathcal{O}_{1})\mathcal{O}_{5}^{2}\bigr)\sqrt{t}
			+\mathcal{O}_{1}\mathcal{O}_{5}t^{3/4}=1+t,\\
			&\mathcal{O}_{1}
			+\mathcal{O}_{1}^{2}\mathcal{O}_{5}t^{1/4}
			+\bigl(\mathcal{O}_{5}^{4}+\mathcal{O}_{1}(2-5\mathcal{O}_{5}^{2})\bigr)\sqrt{t}
			+\mathcal{O}_{1}^{2}\mathcal{O}_{5}t^{3/4}
			+\mathcal{O}_{1}t=0,\\
			&\mathcal{O}_{1}\Bigl(\mathcal{O}_{1}(5+\mathcal{O}_{1})
			+2\mathcal{O}_{5}(-5\mathcal{O}_{1}+\mathcal{O}_{5}^{2})t^{1/4}
			+\mathcal{O}_{1}(5+\mathcal{O}_{1})\sqrt{t}\Bigr)
			=\mathcal{O}_{5}^{2}(1+\sqrt{t}),\\
			&\mathcal{O}_{1}(2+\mathcal{O}_{1})(3+\mathcal{O}_{1})+\mathcal{O}_{5}^{2}
			+4\mathcal{O}_{1}^{2}\mathcal{O}_{5}t^{1/4}\\
			&\quad
			+2\Bigl(\mathcal{O}_{1}(2+\mathcal{O}_{1})(3+\mathcal{O}_{1})
			+\bigl(1+\mathcal{O}_{1}(-15+2(\mathcal{O}_{1}-5)\mathcal{O}_{1})\bigr)\mathcal{O}_{5}^{2}
			-(\mathcal{O}_{1}-5)\mathcal{O}_{1}\mathcal{O}_{5}^{4}\Bigr)\sqrt{t}\\
			&\quad
			+4\mathcal{O}_{1}^{2}\mathcal{O}_{5}t^{3/4}
			+\bigl(\mathcal{O}_{1}(2+\mathcal{O}_{1})(3+\mathcal{O}_{1})+\mathcal{O}_{5}^{2}\bigr)t=0,\\
			&\mathcal{O}_{5}^{6}t^{3/4}
			+\mathcal{O}_{5}(1+\sqrt{t})\bigl(1+(2-22\mathcal{O}_{1})\sqrt{t}+t\bigr)\\
			&\quad
			+\mathcal{O}_{5}^{2}t^{1/4}\Bigl(
			-7+5\mathcal{O}_{1}
			+2\bigl(-7+\mathcal{O}_{1}(70+\mathcal{O}_{1}(110+3\mathcal{O}_{1}))\bigr)\sqrt{t}
			+(-7+5\mathcal{O}_{1})t\Bigr)\\
			&\quad
			=\mathcal{O}_{1}\bigl(26+\mathcal{O}_{1}(35+\mathcal{O}_{1}(22+5\mathcal{O}_{1}))\bigr)
			(1+\sqrt{t})^{2}t^{1/4}
			+2\mathcal{O}_{1}(35+3\mathcal{O}_{1})\mathcal{O}_{5}^{4}t^{3/4}.
		\end{aligned}
	\end{equation}

	\subsection{\texorpdfstring{$\bm{G_{i}}$ or $\bm{G_{p}}$ subgroup}{G_{i} or G_{p} subgroup}}
    
    \label{sec:GiGp}
    
    \(G_{i}\cong \mathbb{Z}_{6}\), \(|G_{i}|=6\), data \(\{0;\{3,1\},\{6,2\}\}\): \\

     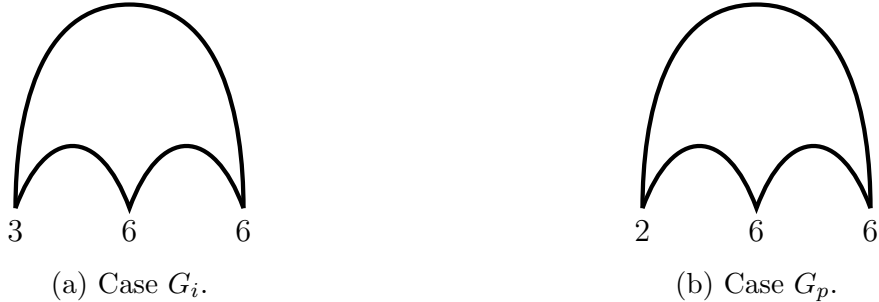
\begin{figure}[h!]
     \begin{subfigure}{0.45\linewidth}
     \centering
	\begin{tikzpicture}
    \draw[ultra thick] (-1.5,-1.5) to[out=90,in=180] (0,1.2) to[out=0,in=90] (1.5,-1.5);
    \draw[ultra thick, looseness=2] (-1.5,-1.5) to[out=70,in=110] (0,-1.5) to[out=70,in=110] (1.5,-1.5);
    \node at (-1.5,-1.8) {$3$};
    \node at (-0,-1.8) {$6$};
    \node at (1.5,-1.8) {$6$}; 
    \end{tikzpicture}
    \caption{Case $G_i$.}
    \end{subfigure}
    \hfill
    \begin{subfigure}{0.45\linewidth}
     \centering
	\begin{tikzpicture}
    \draw[ultra thick] (-1.5,-1.5) to[out=90,in=180] (0,1.2) to[out=0,in=90] (1.5,-1.5);
    \draw[ultra thick, looseness=2] (-1.5,-1.5) to[out=70,in=110] (0,-1.5) to[out=70,in=110] (1.5,-1.5);
    \node at (-1.5,-1.8) {$2$};
    \node at (-0,-1.8) {$6$};
    \node at (1.5,-1.8) {$6$}; 
    \end{tikzpicture}
    \caption{Case $G_p$.}
    \end{subfigure}
    \caption{Branching data of the covering maps in case Section \ref{sec:GiGp}.}
    \end{figure}
	
	\paragraph{$\bm{t=1}$:} $I_{1} \, \cap \, I_{2} \, \cap \, I_{3} \, \cap \, I_{4} \, \cap \, I_{5}$ with $\dim(I_{1}, I_{2}, I_{3}, I_{4}, I_{5}) = (0,0,0,0,0)$
	
\begin{equation}
	I_{1}=\Bigl\langle\, \mathcal{O}-2 \,, \ \forall\,\mathcal{O}\,\Bigr\rangle
	\qquad
		I_{2}
		=
		\Bigl\langle\,
		\mathcal{O}_{i}+1,\;
		\mathcal{O}_{i,i+1}-2,\;
		\mathcal{O}_{i,+1,i+2}+1
		\,\Bigr\rangle .
\end{equation}
\begin{equation}
	I_{3}
	=
	\Bigl\langle\,
	\mathcal{O}_{i},\;
	\mathcal{O}_{i,i+1},\;
	\mathcal{O}_{i,i+1,i+2}+2
	\,\Bigr\rangle , 
	\qquad 
		I_{4}
		=
		\Bigl\langle\,
		\mathcal{O}_{i}+4,\;
		\mathcal{O}_{i,i+1}-8,\;
		\mathcal{O}_{i,i+1,i+2}+10
		\,\Bigr\rangle .
\end{equation}
\begin{equation}
	I_{5}
	=
	\Bigl\langle\,
	\mathcal{O}_{i}+1,\;
	\mathcal{O}_{i,i+1}+1,\;
	\mathcal{O}_{i,i+1,i+2}-2
	\,\Bigr\rangle \qquad i = 1 \dots 6 \,.
\end{equation}

	\paragraph{t-deformed:} $I_{1} \, \cap \, I_{2} \, \cap \, I_{3}$ with $\dim(I_{1}, I_{2}, I_{3}) = (0,0,0)$
	
	\begin{equation}\label{eq:Gi-t-I1}
		I_{1}^{(t)}
		=
		\Bigl\langle
		\mathcal{O}_{i},\;
		\mathcal{O}_{i,i+1},\;
		\mathcal{O}_{i,i+1,i+2}+\frac{1+t}{\sqrt t}
		\Bigr\rangle .
	\end{equation}

	\begin{equation}\label{eq:Gi-t-I2}
		I_{2}^{(t)}:\quad
		\begin{aligned}
			&\mathcal{O}_{i}
			=-\frac{2(1+t^{1/3})}{t^{1/6}},, \quad \mathcal{O}_{i,i+1}
			=2\Bigl(2+t^{-1/3}+t^{1/3}\Bigr),\\[4pt]
			&\mathcal{O}_{i,i+1,i+2}
			=-\frac{1+4t^{1/3}+4t^{2/3}+t}{\sqrt t}.
		\end{aligned}
	\end{equation}

	\begin{equation}\label{eq:Gi-t-I3}
		I_{3}^{(t)}:\quad
		\begin{aligned}
			&\mathcal{O}_{i}
			=t^{-1/6}+t^{1/6},, \quad \mathcal{O}_{i,i+1}
			=\frac12\Bigl(2+(1+i\sqrt3)t^{-1/3}+(1-i\sqrt3)t^{1/3}\Bigr),\\[4pt]
			&\mathcal{O}_{i,i+1,i+2}
			=\frac{(1+t^{1/3})\bigl(1+i\sqrt3+(1-i\sqrt3)t^{2/3}\bigr)}{2\sqrt t}.
		\end{aligned}
	\end{equation}
	
	\nd Also in this case we can identify $G_{i}$ subgroup associated variety with the one emerging from the \(G_{p}\cong \mathbb{Z}_{2}\times\mathbb{Z}_{6}\), \(|G_{p}|=12\), data \(\{0;\{2,1\},\{6,1\},\{6,1\}\}\).

	\subsection{\texorpdfstring{$\bm{G_{k_{1}}}$ subgroup}{G_{k_{1}}subgroup}}
	
	\(G_{k_{1}}\cong \mathbb{Z}_{6}\), \(|G_{k_{1}}|=6\), data \(\{0;\{2,2\},\{3,2\}\}\): \\

     \begin{figure}[h!]
     \centering
	\begin{tikzpicture}
    \draw[ultra thick] (-1.5,-1.5) to[out=90,in=180] (0,1.2) to[out=0,in=90] (1.5,-1.5);
    \draw[ultra thick, looseness=2] (-1.5,-1.5) to[out=70,in=110] (-0.5,-1.5) to[out=70,in=110] (0.5,-1.5) to[out=70,in=110] (1.5,-1.5);
    \node at (-1.5,-1.8) {$2$};
    \node at (-0.5,-1.8) {$2$};
    \node at (0.5,-1.8) {$3$};
    \node at (1.5,-1.8) {$3$}; 
    \end{tikzpicture}
    \caption{Branching data of the covering maps in case $G_{k1}$}
    \end{figure}
	
	\paragraph{$\bm{t=1}$:} $I_{1} \, \cap \, I_{2} \, \cap \, I_{3}$ with $\dim(I_{1}, I_{2}, I_{3}) = (2,0,0)$
	
	\begin{equation}\label{eq:Gk1-I1}
		I_{1}:\qquad
		\begin{aligned}
			&\mathcal{O}_{2}=\mathcal{O}_{4}=\mathcal{O}_{5}=\mathcal{O}_{1}, \quad \mathcal{O}_{1,2}=(\mathcal{O}_{1}-1)\mathcal{O}_{1},\quad
			\mathcal{O}_{2,3}=\mathcal{O}_{6},\quad \mathcal{O}_{4,5}=\mathcal{O}_{1},\quad\\
			&\mathcal{O}_{6,1}=\mathcal{O}_{3}, \quad 
			\mathcal{O}_{3,4}=\mathcal{O}_{1}\mathcal{O}_{3}-\mathcal{O}_{6},
			\mathcal{O}_{5,6}=\mathcal{O}_{1}\mathcal{O}_{6}-\mathcal{O}_{3}, \\
			&\mathcal{O}_{1,2,3}=2-\mathcal{O}_{3}+(\mathcal{O}_{1}-1)\mathcal{O}_{6},\\
			&\mathcal{O}_{2,3,4}=\mathcal{O}_{3}+\mathcal{O}_{6}-2,\quad
			\mathcal{O}_{3,4,5}=2+(\mathcal{O}_{1}-1)\mathcal{O}_{3}-\mathcal{O}_{6},\\[6pt]
			&\mathcal{O}_{1}^{2}-\mathcal{O}_{1}\mathcal{O}_{3}\mathcal{O}_{6}
			+\mathcal{O}_{6}^{2}+(\mathcal{O}_{3}-2)(\mathcal{O}_{3}+\mathcal{O}_{6})=0.
		\end{aligned}
	\end{equation}

	\begin{equation}\label{eq:Gk1-I2}
		I_{2}:\qquad
		\begin{aligned}
			&\mathcal{O}_{1}=\mathcal{O}_{2}=\mathcal{O}_{3}
			=\mathcal{O}_{6}
			=\mathcal{O}_{2,3}
			=\mathcal{O}_{3,4}
			=\mathcal{O}_{5,6}
			=\mathcal{O}_{6,1}
			=\mathcal{O}_{1,2,3}
			=\mathcal{O}_{2,3,4}
			=\mathcal{O}_{3,4,5}=-1,\\
			&\mathcal{O}_{4}=\mathcal{O}_{5}
			=\mathcal{O}_{1,2}
			=\mathcal{O}_{4,5}=2.
		\end{aligned}
	\end{equation}

	\begin{equation}\label{eq:Gk1-I3}
		I_{3}:\qquad
		\begin{aligned}
			&\mathcal{O}_{1}=\mathcal{O}_{2}=\mathcal{O}_{1,2}
			=\mathcal{O}_{3,4}=\mathcal{O}_{5,6}
			=\mathcal{O}_{2,3,4}=2,\\
			&\mathcal{O}_{3}=\mathcal{O}_{4}=\mathcal{O}_{5}=\mathcal{O}_{6}
			=\mathcal{O}_{2,3}=\mathcal{O}_{4,5}=\mathcal{O}_{6,1}
			=\mathcal{O}_{1,2,3}=\mathcal{O}_{3,4,5}=-1.
		\end{aligned}
	\end{equation}

	\paragraph{t-deformed:} $I_{1} \, \cap \, I_{2}$ with $\dim(I_{1}, I_{2}) = (2,0)$
	
	\begin{equation}\label{eq:Gk1-t-I1}
		I_{1}^{(t)}:\quad
		\begin{aligned}
			&\mathcal{O}_{2}=\mathcal{O}_{4}=\mathcal{O}_{5}=\mathcal{O}_{1},\qquad
			\mathcal{O}_{2,3}=\mathcal{O}_{6},\qquad
			\mathcal{O}_{4,5}=\mathcal{O}_{1},\qquad
			\mathcal{O}_{6,1}=\mathcal{O}_{3},\\[4pt]
			&\mathcal{O}_{1,2}=(\mathcal{O}_{1}-1)\mathcal{O}_{1},\qquad
			\mathcal{O}_{3,4}=\mathcal{O}_{1}\mathcal{O}_{3}-\mathcal{O}_{6},\qquad
			\mathcal{O}_{5,6}=-\mathcal{O}_{3}+\mathcal{O}_{1}\mathcal{O}_{6},\\[4pt]
			&\mathcal{O}_{1,2,3}
			=-\mathcal{O}_{3}+(\mathcal{O}_{1}-1)\mathcal{O}_{6}+\frac{1+t}{\sqrt t}, \quad  \mathcal{O}_{2,3,4}=\mathcal{O}_{3}+\mathcal{O}_{6}-\frac{1+t}{\sqrt t},\\
			&\mathcal{O}_{3,4,5}
			=(\mathcal{O}_{1}-1)\mathcal{O}_{3}-\mathcal{O}_{6}+\frac{1+t}{\sqrt t},\\[6pt]
			&\mathcal{O}_{3}+\mathcal{O}_{6}
			-\sqrt t\Bigl(
			\mathcal{O}_{1}^{2}
			-\mathcal{O}_{1}\mathcal{O}_{3}\mathcal{O}_{6}
			+\mathcal{O}_{6}^{2}
			+\mathcal{O}_{3}(\mathcal{O}_{3}+\mathcal{O}_{6})
			\Bigr)
			+(\mathcal{O}_{3}+\mathcal{O}_{6})\,t=0.
		\end{aligned}
	\end{equation}

	\begin{equation}\label{eq:Gk1-t-I2}
		I_{2}^{(t)}:\quad
		\begin{aligned}
			&\mathcal{O}_{3}=\mathcal{O}_{6}=\mathcal{O}_{2,3}
			=\mathcal{O}_{6,1}=\mathcal{O}_{1,2,3}=\mathcal{O}_{3,4,5}
			=t^{-1/6}+t^{1/6},\\[6pt]
			&\mathcal{O}_{1}=\mathcal{O}_{2}
			=\frac12\Bigl(2+(1+i\sqrt3)t^{-1/3}+(1-i\sqrt3)t^{1/3}\Bigr),\\[6pt]
			&\mathcal{O}_{4}=\mathcal{O}_{5}=\mathcal{O}_{4,5}
			=\frac12\Bigl(2+(1-i\sqrt3)t^{-1/3}+(1+i\sqrt3)t^{1/3}\Bigr),\\[6pt]
			&\mathcal{O}_{3,4}=\mathcal{O}_{5,6}=\mathcal{O}_{2,3,4}
			=\frac{(1+t^{1/3})\bigl(1-i\sqrt3+(1+i\sqrt3)t^{2/3}\bigr)}{2\sqrt t},\\[8pt]
			&\mathcal{O}_{1,2}
			=\frac{(1+t^{1/3})\bigl(-1+i\sqrt3+2t^{1/3}+(1-i\sqrt3)t^{2/3}\bigr)}
			{2t^{2/3}}.
		\end{aligned}
	\end{equation}

	\subsection{\texorpdfstring{$\bm{G_{k_{2}}}$ or $\bm{G_{s}}$ subgroup}{G_{k_{2}} or G_{s} subgroup}}

    \label{sec:Gk2Gs}
    
    \(G_{k_{2}}\cong D_{3}\), \(|G_{k_{2}}|=6\), data \(\{0;\{2,2\},\{3,2\}\}\): \\

    \begin{figure}[h!]
    \begin{subfigure}{0.45\linewidth}
     \centering
	\begin{tikzpicture}
    \draw[ultra thick] (-1.5,-1.5) to[out=90,in=180] (0,1.2) to[out=0,in=90] (1.5,-1.5);
    \draw[ultra thick, looseness=2] (-1.5,-1.5) to[out=70,in=110] (-0.5,-1.5) to[out=70,in=110] (0.5,-1.5) to[out=70,in=110] (1.5,-1.5);
    \node at (-1.5,-1.8) {$2$};
    \node at (-0.5,-1.8) {$2$};
    \node at (0.5,-1.8) {$3$};
    \node at (1.5,-1.8) {$3$}; 
    \end{tikzpicture}
     \caption{Case $G_{k2}$.}
    \end{subfigure}
    \hfill
    \begin{subfigure}{0.45\linewidth}
	\begin{tikzpicture}
    \draw[ultra thick] (-1.5,-1.5) to[out=90,in=180] (0,1.2) to[out=0,in=90] (1.5,-1.5);
    \draw[ultra thick, looseness=2] (-1.5,-1.5) to[out=70,in=110] (-0.5,-1.5) to[out=70,in=110] (0.5,-1.5) to[out=70,in=110] (1.5,-1.5);
    \node at (-1.5,-1.8) {$2$};
    \node at (-0.5,-1.8) {$2$};
    \node at (0.5,-1.8) {$2$};
    \node at (1.5,-1.8) {$3$}; 
    \end{tikzpicture}
     \caption{Case $G_{s}$.}
    \end{subfigure}
    \caption{Branching data of the covering maps in case Section \ref{sec:Gk2Gs}.}
    \end{figure}
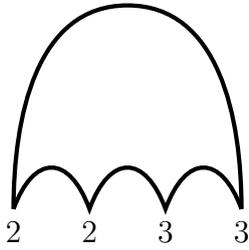
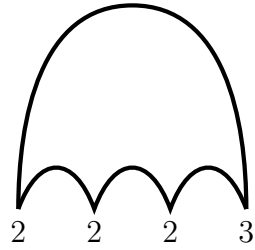
	
	\paragraph{$\bm{t=1}$:} $I_{1}$ with $\dim(I_{1}) = 2$
	\begin{equation}\label{eq:Gk2-I1}
		I_{1}:\qquad
		\begin{aligned}
			&\mathcal{O}_{2}=\mathcal{O}_{4}=\mathcal{O}_{5}=\mathcal{O}_{1},\qquad
			\mathcal{O}_{1,2}=\mathcal{O}_{4,5}=(\mathcal{O}_{1}-1)\mathcal{O}_{1},\\
			&\mathcal{O}_{5,6}=\mathcal{O}_{2,3},\qquad
			\mathcal{O}_{2,3,4}=\mathcal{O}_{3},\\
			&\mathcal{O}_{3,4}=\mathcal{O}_{6,1}=2+(\mathcal{O}_{1}-1)\mathcal{O}_{3}-\mathcal{O}_{2,3},\\
			&\mathcal{O}_{1,2,3}=2-\mathcal{O}_{3}+(\mathcal{O}_{1}-1)\mathcal{O}_{2,3},\\
			&\mathcal{O}_{3,4,5}=\mathcal{O}_{1}\bigl(2+(\mathcal{O}_{1}-2)\mathcal{O}_{3}-\mathcal{O}_{2,3}\bigr)
			+\mathcal{O}_{2,3},\\[6pt]
			&\mathcal{O}_{1}^{2}-\mathcal{O}_{1}\mathcal{O}_{3}\mathcal{O}_{2,3}
			+\mathcal{O}_{2,3}^{2}+(\mathcal{O}_{3}-2)(\mathcal{O}_{3}+\mathcal{O}_{2,3})=0.
		\end{aligned}
	\end{equation}

	\paragraph{t-deformed:} $I_{1}$ with $\dim(I_{1}) = 2$
	
	\begin{equation}\label{eq:Gk2-t-I1}
		I_{1}^{(t)}:\quad
		\begin{aligned}
			&\mathcal{O}_{2}=\mathcal{O}_{4}=\mathcal{O}_{5}=\mathcal{O}_{1},\quad
			\mathcal{O}_{5,6}=\mathcal{O}_{2,3},\quad
			\mathcal{O}_{2,3,4}=\mathcal{O}_{3},\quad \mathcal{O}_{1,2}=\mathcal{O}_{4,5}=(\mathcal{O}_{1}-1)\mathcal{O}_{1},\\
			&\mathcal{O}_{3,4}=\mathcal{O}_{6,1}
			=(\mathcal{O}_{1}-1)\mathcal{O}_{3}-\mathcal{O}_{2,3}+\frac{1+t}{\sqrt t}, \quad \mathcal{O}_{1,2,3}
			=-\mathcal{O}_{3}+(\mathcal{O}_{1}-1)\mathcal{O}_{2,3}+\frac{1+t}{\sqrt t},\\
			&\mathcal{O}_{3,4,5}
			=\mathcal{O}_{2,3}
			+\mathcal{O}_{1}\Bigl((\mathcal{O}_{1}-2)\mathcal{O}_{3}-\mathcal{O}_{2,3}
			+\frac{1+t}{\sqrt t}\Bigr),\\[6pt]
			&\mathcal{O}_{3}+\mathcal{O}_{2,3}
			-\sqrt t\Bigl(
			\mathcal{O}_{1}^{2}
			-\mathcal{O}_{1}\mathcal{O}_{3}\mathcal{O}_{2,3}
			+\mathcal{O}_{2,3}^{2}
			+\mathcal{O}_{3}(\mathcal{O}_{3}+\mathcal{O}_{2,3})
			\Bigr)
			+(\mathcal{O}_{3}+\mathcal{O}_{2,3})t=0 \,,
		\end{aligned}
	\end{equation} 
	where $G_{k2}$ case appears to be equivalent to $ G_{s} \cong D_{6}$, \(|G_{s}|=12\), data \(\{0;\{2,3\},\{3,1\}\}\).

	\subsection{\texorpdfstring{$\bm{G_{l}}$ subgroup}{G_{l} subgroup}}
	
	\(G_{l}\cong \mathbb{Z}_{8}\), \(|G_{l}|=8\), data \(\{0;\{2,1\},\{8,2\},\{8,1\}\}\):\\

     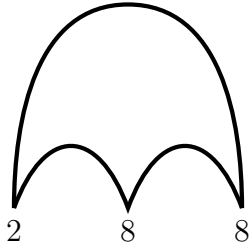
\begin{figure}[h!]
     \centering
	\begin{tikzpicture}
    \draw[ultra thick] (-1.5,-1.5) to[out=90,in=180] (0,1.2) to[out=0,in=90] (1.5,-1.5);
    \draw[ultra thick, looseness=2] (-1.5,-1.5) to[out=70,in=110] (0,-1.5) to[out=70,in=110] (1.5,-1.5);
    \node at (-1.5,-1.8) {$2$};
    \node at (0,-1.8) {$8$};
    \node at (1.5,-1.8) {$8$};
    \end{tikzpicture}
    \caption{Branching data of the covering maps in case $G_{l}$}
    \end{figure}
	
	\paragraph{$\bm{t=1}$:} $I_{1} \, \cap \, I_{2} \, \cap \, I_{3} \, \cap \, I_{4}$ with $\dim(I_{1}, I_{2}, I_{3}, I_{4}) = (0,0,0,0)$
	
	\begin{equation}\label{eq:Gl-I1}
		I_{1}
		=
		\Bigl\langle\,
		\mathcal{O}-2,\ \forall\,\mathcal{O}
		\,\Bigr\rangle .
	\end{equation}

	\begin{equation}\label{eq:Gl-I2}
		I_{2}:\qquad
		\begin{aligned}
			&\mathcal{O}_{1}=\mathcal{O}_{5}=\mathcal{O}_{1,2}
			=\mathcal{O}_{4,5}=\mathcal{O}_{5,6}
			=\mathcal{O}_{6,1}=\mathcal{O}_{1,2,3}
			=\mathcal{O}_{3,4,5}=-2,\\
			&\mathcal{O}_{2}=\mathcal{O}_{3}=\mathcal{O}_{4}
			=\mathcal{O}_{6}=\mathcal{O}_{2,3}
			=\mathcal{O}_{3,4}
			=\mathcal{O}_{2,3,4}=2.
		\end{aligned}
	\end{equation} 
	
	\begin{equation}\label{eq:Gl-I3}
		I_{3}:\qquad
		\begin{aligned}
			&\mathcal{O}_{1}=\mathcal{O}_{2}=\mathcal{O}_{3}=\mathcal{O}_{4}
			=2-2\sqrt{2},\\
			&\mathcal{O}_{5}=\mathcal{O}_{4,5}=\mathcal{O}_{6,1}
			=\mathcal{O}_{3,4,5}=2(\sqrt{2}-1),\\
			&\mathcal{O}_{6}=\mathcal{O}_{2,3}=\mathcal{O}_{3,4}
			=6-4\sqrt{2},\\
			&\mathcal{O}_{1,2}=\mathcal{O}_{2,3,4}=10-6\sqrt{2},\\
			&\mathcal{O}_{5,6}=26-18\sqrt{2},\qquad
			\mathcal{O}_{1,2,3}=18-14\sqrt{2}.
		\end{aligned}
	\end{equation}

	\begin{equation}\label{eq:Gl-I4}
		I_{4}:\qquad
		\begin{aligned} 
			&\mathcal{O}_{1}=-2(1+\sqrt{2}), \quad \mathcal{O}_{6}=-6-4\sqrt{2}, \quad \mathcal{O}_{2,3}=\mathcal{O}_{3,4}=6+4\sqrt{2},\\
			&\mathcal{O}_{2}=\mathcal{O}_{3}=\mathcal{O}_{4}
			=\mathcal{O}_{5}=\mathcal{O}_{4,5}=\mathcal{O}_{6,1}
			=\mathcal{O}_{3,4,5}=2(1+\sqrt{2}),\\
			&\mathcal{O}_{1,2}=-2(5+3\sqrt{2}),\qquad
			\mathcal{O}_{5,6}=-2(13+9\sqrt{2}),\\
			&\mathcal{O}_{1,2,3}=-2(9+7\sqrt{2}),\qquad
			\mathcal{O}_{2,3,4}=10+6\sqrt{2}.
		\end{aligned}
	\end{equation}

	\paragraph{t-deformed:} $I_{1} \, \cap \, I_{2} \, \cap \, I_{3} \, \cap \, I_{4} \, \cap \, I_{5}$ with $\dim(I_{1}, I_{2}, I_{3}, I_{4}, I_{5}) = (0,0,0,0,0)$
	
	\begin{equation}\label{eq:Gl-t-I1}
		I_{1}^{(t)}:\qquad
		\begin{aligned}
			&\mathcal{O}_{1}=\mathcal{O}_{2}=\mathcal{O}_{3}=\mathcal{O}_{4}
			=-\mathcal{O}_{5}=-\mathcal{O}_{4,5}=-\mathcal{O}_{6,1}=-\mathcal{O}_{3,4,5}
			=2-\frac{\sqrt{2}\,(1+\sqrt{t})}{t^{1/4}},\\[6pt]
			&\mathcal{O}_{2,3}=\mathcal{O}_{3,4}=-\mathcal{O}_{6}
			=4+\frac{1}{\sqrt{t}}
			-\frac{2\sqrt{2}\,(1+\sqrt{t})}{t^{1/4}}
			+\sqrt{t},\\[6pt]
			&\mathcal{O}_{1,2}=\mathcal{O}_{2,3,4}
			=6+\frac{2}{\sqrt{t}}
			-\frac{3\sqrt{2}\,(1+\sqrt{t})}{t^{1/4}}
			+2\sqrt{t},\\[6pt]
			&\mathcal{O}_{5,6}
			=14-\frac{\sqrt{2}\,(1+\sqrt{t})^{3}}{t^{3/4}}
			+\frac{6}{\sqrt{t}}
			-\frac{5\sqrt{2}\,(1+\sqrt{t})}{t^{1/4}}
			+6\sqrt{t},\\[6pt]
			&\mathcal{O}_{1,2,3}
			=10-\frac{\sqrt{2}\,(1+\sqrt{t})^{3}}{t^{3/4}}
			+\frac{4}{\sqrt{t}}
			-\frac{3\sqrt{2}\,(1+\sqrt{t})}{t^{1/4}}
			+4\sqrt{t}.
		\end{aligned}
	\end{equation}
	
	\begin{equation}\label{eq:Gl-t-I2}
		I_{2}^{(t)}:\qquad
		\begin{aligned}
			&\mathcal{O}_{1}=\mathcal{O}_{2}=\mathcal{O}_{3}=\mathcal{O}_{4}
			=\mathcal{O}_{5}=\mathcal{O}_{4,5}=\mathcal{O}_{6,1}=\mathcal{O}_{3,4,5}
			=2-\frac{i\sqrt{2}\,(\sqrt{t}-1)}{t^{1/4}},\\[6pt]
			&\mathcal{O}_{6}=\mathcal{O}_{2,3}=\mathcal{O}_{3,4}
			=4-\frac{1}{\sqrt{t}}
			-\frac{2i\sqrt{2}\,(\sqrt{t}-1)}{t^{1/4}}
			-\sqrt{t},\\[6pt]
			&\mathcal{O}_{1,2}=\mathcal{O}_{2,3,4}
			=6-\frac{2}{\sqrt{t}}
			-\frac{3i\sqrt{2}\,(\sqrt{t}-1)}{t^{1/4}}
			-2\sqrt{t},\\[6pt]
			&\mathcal{O}_{5,6}
			=14+\frac{i\sqrt{2}\,(\sqrt{t}-1)^{3}}{t^{3/4}}
			-\frac{6}{\sqrt{t}}
			-\frac{5i\sqrt{2}\,(\sqrt{t}-1)}{t^{1/4}}
			-6\sqrt{t},\\[6pt]
			&\mathcal{O}_{1,2,3}
			=10+\frac{i\sqrt{2}\,(\sqrt{t}-1)^{3}}{t^{3/4}}
			-\frac{4}{\sqrt{t}}
			-\frac{3i\sqrt{2}\,(\sqrt{t}-1)}{t^{1/4}}
			-4\sqrt{t}.
		\end{aligned}
	\end{equation}

	\begin{equation}\label{eq:Gl-t-I3}
		I_{3}^{(t)}:\quad
		\begin{aligned}
			&\mathcal{O}_{2}=\mathcal{O}_{3}=\mathcal{O}_{4}
			=\mathcal{O}_{5}=\mathcal{O}_{4,5}=\mathcal{O}_{6,1}=\mathcal{O}_{3,4,5}=-\mathcal{O}_{1}
			=2+\frac{\sqrt{2}\,(1+\sqrt{t})}{t^{1/4}},\\[6pt]
			&\mathcal{O}_{2,3}=\mathcal{O}_{3,4}
			=4+\frac{1}{\sqrt{t}}
			+\frac{2\sqrt{2}\,(1+\sqrt{t})}{t^{1/4}}
			+\sqrt{t},\\[6pt]
			&\mathcal{O}_{2,3,4}
			=6+\frac{2}{\sqrt{t}}
			+\frac{3\sqrt{2}\,(1+\sqrt{t})}{t^{1/4}}
			+2\sqrt{t},\\[6pt]
			&\mathcal{O}_{6}
			=-4-\frac{1+2\sqrt{2}\,t^{1/4}+2\sqrt{2}\,t^{3/4}+t}{\sqrt{t}},\\[6pt]
			&\mathcal{O}_{1,2}
			=-2(3+\sqrt{t})-\frac{2}{\sqrt{t}}
			-\frac{3\sqrt{2}\,(1+\sqrt{t})}{t^{1/4}},\\[6pt]
			&\mathcal{O}_{5,6}
			=-14-\frac{\sqrt{2}+6t^{1/4}+8\sqrt{2}\sqrt{t}+8\sqrt{2}t+6t^{5/4}
				+\sqrt{2}t^{3/2}}{t^{3/4}},\\[6pt]
			&\mathcal{O}_{1,2,3}
			=-10-\frac{\sqrt{2}+4t^{1/4}+6\sqrt{2}\sqrt{t}+6\sqrt{2}t+4t^{5/4}
				+\sqrt{2}t^{3/2}}{t^{3/4}}.
		\end{aligned}
	\end{equation}

	\begin{equation}\label{eq:Gl-t-I4}
		I_{4}^{(t)}:\qquad
		\begin{aligned}
			&\mathcal{O}_{2}=\mathcal{O}_{3}=\mathcal{O}_{4}
			=-\mathcal{O}_{5}=-\mathcal{O}_{1}=-\mathcal{O}_{4,5}=-\mathcal{O}_{6,1}=-\mathcal{O}_{3,4,5}
			=2+\frac{i\sqrt{2}\,(\sqrt{t}-1)}{t^{1/4}},\\[6pt]
			&\mathcal{O}_{6}=\mathcal{O}_{2,3}=\mathcal{O}_{3,4}
			=4-\frac{1}{\sqrt{t}}
			+\frac{2i\sqrt{2}\,(\sqrt{t}-1)}{t^{1/4}}
			-\sqrt{t},\\[6pt]
			&-\mathcal{O}_{1,2}=\mathcal{O}_{2,3,4}
			=6-\frac{2}{\sqrt{t}}
			+\frac{3i\sqrt{2}\,(\sqrt{t}-1)}{t^{1/4}}
			-2\sqrt{t},\\[6pt]
			&\mathcal{O}_{5,6}
			=-14+\frac{i\sqrt{2}\,(\sqrt{t}-1)^{3}}{t^{3/4}}
			+\frac{6}{\sqrt{t}}
			-\frac{5i\sqrt{2}\,(\sqrt{t}-1)}{t^{1/4}}
			+6\sqrt{t},\\[6pt]
			&\mathcal{O}_{1,2,3}
			=-10+\frac{i\sqrt{2}\,(\sqrt{t}-1)^{3}}{t^{3/4}}
			+\frac{4}{\sqrt{t}}
			-\frac{3i\sqrt{2}\,(\sqrt{t}-1)}{t^{1/4}}
			+4\sqrt{t}.
		\end{aligned}
	\end{equation}

	\begin{equation}\label{eq:Gl-t-I5}
		I_{5}^{(t)}:\quad
		\begin{aligned}
			&\mathcal{O}_{2}=\mathcal{O}_{3}=\mathcal{O}_{4}=2, \quad \mathcal{O}_{1}=\mathcal{O}_{1,2}
			=-\frac{(1+i)(\sqrt{t}-i)}{t^{1/4}},\\[4pt]
			&\mathcal{O}_{5}=\mathcal{O}_{4,5}=\mathcal{O}_{6,1}=\mathcal{O}_{3,4,5}
			=-\frac{(1-i)(\sqrt{t}+i)}{t^{1/4}},\\[4pt]
			&\mathcal{O}_{6}=\frac{1+t}{\sqrt{t}}, \quad \mathcal{O}_{2,3}=\mathcal{O}_{3,4}
			=2-\frac{i(t-1)}{\sqrt{t}},\\[4pt]
			&\mathcal{O}_{5,6}=\mathcal{O}_{1,2,3}
			=-\frac{(1-i)(i+t^{3/2})}{t^{3/4}}, \quad \mathcal{O}_{2,3,4}
			=2-\frac{2i(t-1)}{\sqrt{t}}.
		\end{aligned}
	\end{equation}

	\subsection{\texorpdfstring{$\bm{G_{m}}$ subgroup}{G_{m} subgroup}}

	\(G_{m}\cong \widetilde{D}_{2}\), \(|G_{m}|=8\), data \(\{0;\{4,1\},\{4,1\},\{4,1\}\}\): 

     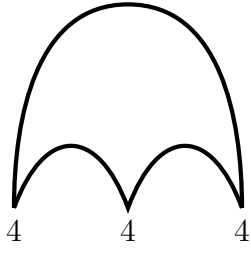
\begin{figure}[h!]
     \centering
	\begin{tikzpicture}
    \draw[ultra thick] (-1.5,-1.5) to[out=90,in=180] (0,1.2) to[out=0,in=90] (1.5,-1.5);
    \draw[ultra thick, looseness=2] (-1.5,-1.5) to[out=70,in=110] (0,-1.5) to[out=70,in=110] (1.5,-1.5);
    \node at (-1.5,-1.8) {$4$};
    \node at (0,-1.8) {$4$};
    \node at (1.5,-1.8) {$4$};
    \end{tikzpicture}
    \caption{Branching data of the covering maps in case $G_{m}$}
    \end{figure}
	
	\paragraph{$\bm{t=1}$:} $I_{1} \, \cap \, I_{2} \, \cap \, I_{3} \, \cap \, I_{4}$ with $\dim(I_{1}, I_{2}, I_{3}, I_{4}) = (0,0,0,0)$
	\begin{equation}\label{eq:Gm-I1}
		I_{1}
		=
		\Bigl\langle\,
		\mathcal{O}-2,\ \forall\,\mathcal{O}
		\,\Bigr\rangle .
	\end{equation} 
	\begin{equation}\label{eq:Gm-I2}
		I_{2}:\quad
		\begin{aligned}
			&\mathcal{O}_{1}=\mathcal{O}_{5}=\mathcal{O}_{1,2}
			=\mathcal{O}_{4,5}=\mathcal{O}_{5,6}
			=\mathcal{O}_{6,1}=\mathcal{O}_{1,2,3}
			=\mathcal{O}_{3,4,5}=-2,\\[4pt]
			&\mathcal{O}_{2}=\mathcal{O}_{3}=\mathcal{O}_{4}=\mathcal{O}_{6}
			=\mathcal{O}_{2,3}=\mathcal{O}_{3,4}
			=\mathcal{O}_{2,3,4}=2 .
		\end{aligned}
	\end{equation}
	\begin{equation}\label{eq:Gm-I3}
		I_{3}:\quad
		\begin{aligned}
			&\mathcal{O}_{1}=\mathcal{O}_{6}=\mathcal{O}_{2,3}=\mathcal{O}_{3,4}
			=\mathcal{O}_{4,5}=\mathcal{O}_{6,1}=\mathcal{O}_{1,2,3}=2,\\
			&\mathcal{O}_{2}=\mathcal{O}_{3}=\mathcal{O}_{4}=\mathcal{O}_{5}
			=\mathcal{O}_{1,2}=\mathcal{O}_{5,6}
			=\mathcal{O}_{2,3,4}=\mathcal{O}_{3,4,5}=-2.
		\end{aligned}
	\end{equation}
	\begin{equation}\label{eq:Gm-I4}
		I_{4}:\quad
		\begin{aligned}
			&\mathcal{O}_{1}=\mathcal{O}_{2}=\mathcal{O}_{3}=\mathcal{O}_{4}
			=\mathcal{O}_{4,5}=\mathcal{O}_{6,1}
			=\mathcal{O}_{1,2,3}=\mathcal{O}_{2,3,4}=-2,\\[4pt]
			&\mathcal{O}_{5}=\mathcal{O}_{6}=\mathcal{O}_{1,2}
			=\mathcal{O}_{2,3}=\mathcal{O}_{3,4}
			=\mathcal{O}_{5,6}=\mathcal{O}_{3,4,5}=2.
		\end{aligned}
	\end{equation}

    \paragraph{t-deformed:}
	$I^{(t)}_{1} \,\cap\, I^{(t)}_{2}\,\cap\, I^{(t)}_{3}\,\cap\, I^{(t)}_{4}$ with $\dim(I^{(t)}_{1},I^{(t)}_{2},I^{(t)}_{3},I^{(t)}_{4})=(0,0,0,0)$ 

    \begin{equation}\label{eq:Gm-t-I1}
    I_{1}^{(t)}:\quad
    \begin{aligned}
    &\mathcal{O}_{1}=\mathcal{O}_{4,5}=\mathcal{O}_{6,1}
    =2-\frac{i\sqrt{2}\,(\sqrt{t}-1)}{t^{1/4}},\\[4pt]
    &\mathcal{O}_{2}=\mathcal{O}_{3}=\mathcal{O}_{4}=\mathcal{O}_{5}
    =\mathcal{O}_{3,4,5}
    =-2+\frac{i\sqrt{2}\,(\sqrt{t}-1)}{t^{1/4}},\\[4pt]
    &\mathcal{O}_{6}=\mathcal{O}_{2,3}=\mathcal{O}_{3,4}
    =4-\frac{1}{\sqrt{t}}
    -\frac{2i\sqrt{2}\,(\sqrt{t}-1)}{t^{1/4}}
    -\sqrt{t},\\[4pt]
    &\mathcal{O}_{1,2}=\mathcal{O}_{2,3,4}
    =-6+\frac{2}{\sqrt{t}}
    +\frac{3i\sqrt{2}\,(\sqrt{t}-1)}{t^{1/4}}
    +2\sqrt{t},\\[4pt]
    &\mathcal{O}_{5,6}
    =-14-\frac{i\sqrt{2}\,(\sqrt{t}-1)^{3}}{t^{3/4}}
    +\frac{6}{\sqrt{t}}
    +\frac{5i\sqrt{2}\,(\sqrt{t}-1)}{t^{1/4}}
    +6\sqrt{t},\\[4pt]
    &\mathcal{O}_{1,2,3}
    =10+\frac{i\sqrt{2}\,(\sqrt{t}-1)^{3}}{t^{3/4}}
    -\frac{4}{\sqrt{t}}
    -\frac{3i\sqrt{2}\,(\sqrt{t}-1)}{t^{1/4}}
    -4\sqrt{t}.
    \end{aligned}
    \end{equation}

    \begin{equation}\label{eq:Gm-t-I2}
    I_{2}^{(t)}:\quad
    \begin{aligned}
    &\mathcal{O}_{1}=\mathcal{O}_{2}=\mathcal{O}_{3}=\mathcal{O}_{4}
    =\mathcal{O}_{5}=\mathcal{O}_{4,5}=\mathcal{O}_{6,1}=\mathcal{O}_{3,4,5}
    =2-\frac{i\sqrt{2}\,(\sqrt{t}-1)}{t^{1/4}},\\[4pt]
    &\mathcal{O}_{6}=\mathcal{O}_{2,3}=\mathcal{O}_{3,4}
    =4-\frac{1}{\sqrt{t}}
    -\frac{2i\sqrt{2}\,(\sqrt{t}-1)}{t^{1/4}}
    -\sqrt{t},\\[4pt]
    &\mathcal{O}_{1,2}=\mathcal{O}_{2,3,4}
    =6-\frac{2}{\sqrt{t}}
    -\frac{3i\sqrt{2}\,(\sqrt{t}-1)}{t^{1/4}}
    -2\sqrt{t},\\[4pt]
    &\mathcal{O}_{5,6}
    =14+\frac{i\sqrt{2}\,(\sqrt{t}-1)^{3}}{t^{3/4}}
    -\frac{6}{\sqrt{t}}
    -\frac{5i\sqrt{2}\,(\sqrt{t}-1)}{t^{1/4}}
    -6\sqrt{t},\\[4pt]
    &\mathcal{O}_{1,2,3}
    =10+\frac{i\sqrt{2}\,(\sqrt{t}-1)^{3}}{t^{3/4}}
    -\frac{4}{\sqrt{t}}
    -\frac{3i\sqrt{2}\,(\sqrt{t}-1)}{t^{1/4}}
    -4\sqrt{t}.
    \end{aligned}
    \end{equation}

	\subsection{\texorpdfstring{$\bm{G_{n}}$ subgroup}{G_{n} subgroup}}
	
	\(G_{n}\cong D_{4}\), \(|G_{n}|=8\), data \(\{0;\{2,3\},\{4,1\}\}\).
	
	\paragraph{$\bm{t=1}$:}
	$I_{1} \,\cap\, I_{2} \,\cap\, I_{3}$ with $\dim(I_{1},I_{2},I_{3})=(2,0,0)$.
	
	\begin{equation}\label{eq:Gn-I1}
		I_{1}:\quad
		\begin{aligned}
			&\mathcal{O}_{4}=\mathcal{O}_{1},\quad
			\mathcal{O}_{4,5}=\mathcal{O}_{1,2},\quad
			\mathcal{O}_{6,1}=\mathcal{O}_{3,4}, \quad  \mathcal{O}_{3,4,5}=2(\mathcal{O}_{1}-1),\\[4pt]
			&\mathcal{O}_{2}=\mathcal{O}_{3}=\mathcal{O}_{5}=\mathcal{O}_{6}
			=\frac{\mathcal{O}_{1,2}+\mathcal{O}_{3,4}}{\mathcal{O}_{1}},\\[4pt]
			&\mathcal{O}_{2,3}=\mathcal{O}_{5,6}
			=\mathcal{O}_{1}(\mathcal{O}_{1}-2)+\frac{\mathcal{O}_{1,2}\mathcal{O}_{3,4}}{2},\\[4pt]
			&\mathcal{O}_{1,2,3}
			=2-\mathcal{O}_{1}+\frac{\mathcal{O}_{1,2}(\mathcal{O}_{1,2}+\mathcal{O}_{3,4})}{2\mathcal{O}_{1}},\\
			&\mathcal{O}_{2,3,4}
			=2-\mathcal{O}_{1}+\frac{\mathcal{O}_{3,4}(\mathcal{O}_{1,2}+\mathcal{O}_{3,4})}{2\mathcal{O}_{1}},\\
			&2\mathcal{O}_{1}^{3}(\mathcal{O}_{1}-2)-\mathcal{O}_{1}^{2}\mathcal{O}_{1,2}\mathcal{O}_{3,4}
			+(\mathcal{O}_{1,2}+\mathcal{O}_{3,4})^{2}=0.
		\end{aligned}
	\end{equation}

	\begin{equation}\label{eq:Gn-I2}
		I_{2}
		=
		\Bigl\langle
		\mathcal{O}_{i},\;
		\mathcal{O}_{i,i+1},\;
		\mathcal{O}_{i,i+1,i+2}+2
		\Bigr\rangle \qquad i=1,\dots,6.
	\end{equation} 
	\begin{equation}\label{eq:Gn-I3}
		I_{3}
		=
		\Bigl\langle
		\mathcal{O}_{1}+2,\;
		\mathcal{O}_{4}+2,\;
		\mathcal{O}_{3,4,5}-2,\;
		\bar{\mathcal{O}}
		\Bigr\rangle \,,
	\end{equation}
	where bar implies complementary to the preceding three conditions.
%
	
	\paragraph{t-deformed:}
	$I^{(t)}_{1} \,\cap\, I^{(t)}_{2}\,\cap\, I^{(t)}_{3}$ with $\dim(I^{(t)}_{1},I^{(t)}_{2},I^{(t)}_{3})=(2,0,0)$.
	
	\begin{equation}\label{eq:Gn-t-I1}
		I^{(t)}_{1}:\quad
		\begin{aligned}
			&\mathcal{O}_{3}=\mathcal{O}_{2},\qquad
			\mathcal{O}_{5}=\mathcal{O}_{2},\qquad
			\mathcal{O}_{6}=\mathcal{O}_{2},\qquad
			\mathcal{O}_{4,5}=\mathcal{O}_{1,2},\qquad
			\mathcal{O}_{6,1}=\mathcal{O}_{3,4},\\[4pt]
			&\mathcal{O}_{5,6}=\mathcal{O}_{2,3}, \quad \mathcal{O}_{1}=\mathcal{O}_{4}
			=\frac{\mathcal{O}_{1,2}+\mathcal{O}_{3,4}}{\mathcal{O}_{2}},\\[4pt]
			&2\mathcal{O}_{2}^{2}\mathcal{O}_{1,2,3}
			-\mathcal{O}_{2}^{3}\mathcal{O}_{1,2}
			+2\mathcal{O}_{1,2}+2\mathcal{O}_{3,4}
			-2\mathcal{O}_{2}^{2}\frac{1+t}{\sqrt{t}}=0,\\[4pt]
			&2\mathcal{O}_{2}^{2}\mathcal{O}_{2,3,4}
			-\mathcal{O}_{2}^{3}\mathcal{O}_{3,4}
			+2\mathcal{O}_{1,2}+2\mathcal{O}_{3,4}
			-2\mathcal{O}_{2}^{2}\frac{1+t}{\sqrt{t}}=0,\\[4pt]
			&\mathcal{O}_{2}\mathcal{O}_{3,4,5}
			-2(\mathcal{O}_{1,2}+\mathcal{O}_{3,4})
			+\mathcal{O}_{2}\frac{1+t}{\sqrt{t}}=0,\\[4pt]
			&2\mathcal{O}_{2}^{2}\mathcal{O}_{2,3}
			-\mathcal{O}_{2}^{2}\mathcal{O}_{1,2}\mathcal{O}_{3,4}
			+2(\mathcal{O}_{1,2}+\mathcal{O}_{3,4})^{2}
			-2\mathcal{O}_{2}(\mathcal{O}_{1,2}+\mathcal{O}_{3,4})\frac{1+t}{\sqrt{t}}=0,\\
			&\mathcal{O}_{2}^{4}-\mathcal{O}_{2}^{2}\mathcal{O}_{1,2}\mathcal{O}_{3,4}
			+2(\mathcal{O}_{1,2}+\mathcal{O}_{3,4})^{2}
			-2\mathcal{O}_{2}(\mathcal{O}_{1,2}+\mathcal{O}_{3,4})\frac{1+t}{\sqrt{t}}=0.
		\end{aligned}
	\end{equation}
	
	\begin{equation}\label{eq:Gn-t-I2}
		I^{(t)}_{2}
		=
		\Bigl\langle
		\mathcal{O}_{i},\;
		\mathcal{O}_{i,i+1},\;
		\mathcal{O}_{i,i+1,i+2}+\frac{1+t}{\sqrt t}
		\Bigr\rangle .
	\end{equation}
	\begin{equation}\label{eq:Gn-t-I3}
		I^{(t)}_{3}
		=
		\Bigl\langle
		\mathcal{O}_{1}+\frac{1+t}{\sqrt t},\;
		\mathcal{O}_{4}+\frac{1+t}{\sqrt t},\;
		\mathcal{O}_{3,4,5}-\frac{1+t}{\sqrt t},\;
		\bar{\mathcal{O}}
		\Bigr\rangle .
	\end{equation}

    \subsection{\texorpdfstring{$\bm{G_{r}}$ or $\bm{G_{w}}$ subgroup}{G_{r} or G_{w} subgroup}}

	\(G_{r}\cong D_{4}\), \(|G_{r}|=12\), data \(\{0;\{3,1\},\{4,2\}\}\).
	
	\paragraph{$\bm{t=1}$:}
	$I_{1} \,\cap\, I_{2} \,\cap\, I_{3}$ with $\dim(I_{1},I_{2},I_{3})=(0,0,0)$.
	
	\begin{equation}\label{eq:Gr-I1}
		I_{1}
		=
		\Bigl\langle
		\mathcal{O}_{i},\;
		\mathcal{O}_{i,i+1},\;
		\mathcal{O}_{i,i+1,i+2}+2
		\Bigr\rangle \,, \quad 
		I_{2}
		=
		\Bigl\langle\,
		\mathcal{O}-2,\ \forall\,\mathcal{O}
		\,\Bigr\rangle
	\end{equation}

	\begin{equation}\label{eq:Gr-I3}
		I_{3}
		=
		\Bigl\langle
		\mathcal{O}_{i}+4,\;
		\mathcal{O}_{i,i+1}-8,\;
		\mathcal{O}_{i,i+1,i+2}+10
		\Bigr\rangle .
	\end{equation}
	
	\paragraph{t-deformed:}
	$I^{(t)}_{1} \,\cap\, I^{(t)}_{2}$ with $\dim(I^{(t)}_{1},I^{(t)}_{2})=(0,0)$.
	
	\begin{equation}\label{eq:Gr-t-I1}
		I_{1}^{(t)}
		=
		\Bigl\langle
		\mathcal{O}_{i},\;
		\mathcal{O}_{i,i+1},\;
		\mathcal{O}_{i,i+1,i+2}+\frac{1+t}{\sqrt t}
		\Bigr\rangle .
	\end{equation}
	
	\begin{equation}\label{eq:Gr-t-I2}
		I_{2}^{(t)}:\quad
		\begin{aligned}
			&\mathcal{O}_{i}
			=-\frac{2(-1)^{2/3}\bigl((-1)^{2/3}+t^{1/3}\bigr)}{t^{1/6}},\\[4pt]
			&\mathcal{O}_{i,i+1}
			=4+\frac{2(-1)^{2/3}}{t^{1/3}}-2(-1)^{1/3}t^{1/3},\\[4pt]
			&\mathcal{O}_{i,i+1,i+2}
			=-\frac{\bigl(1-2(-1)^{1/3}t^{1/3}\bigr)^{2}+t}{\sqrt t}.
		\end{aligned}
	\end{equation}
	
	\nd The $G_{r}$ associated variety becomes equal to the case of the \(G_{w}\cong \mathbb{Z}_{2}\rtimes(\mathbb{Z}_{2}\times\mathbb{Z}_{2}\times\mathbb{Z}_{3})\) \tx{subgroup}, where \(|G_{w}|=24\) and ramification data is \(\{0;\{2,1\},\{4,1\},\{6,1\}\}\).

	\subsection{\texorpdfstring{$\bm{G_{u}}$ subgroup}{G_{u} subgroup}}
	
	\(G_{u}\cong D_{2}\), \(|G_{u}|=16\), data \(\{0;\{2,1\},\{4,1\},\{8,1\}\}\).
	
	\paragraph{$\bm{t=1}$:}
	$I_{1} \,\cap\, I_{2}$ with $\dim(I_{1},I_{2})=(0,0)$.
	
	\begin{equation}\label{eq:Gu-I1}
		I_{1}
		=
		\Bigl\langle
		\mathcal{O}-2,\ \forall\,\mathcal{O}
		\Bigr\rangle .
	\end{equation}
	
	\begin{equation}\label{eq:Gu-I2}
		I_{2}:\quad
		\begin{aligned}
			&\mathcal{O}_{1}=\mathcal{O}_{4}=\mathcal{O}_{2,3}=\mathcal{O}_{5,6}
			=\mathcal{O}_{1,2,3}=\mathcal{O}_{2,3,4}=\mathcal{O}_{3,4,5}=2,\\
			&\mathcal{O}_{2}=\mathcal{O}_{3}=\mathcal{O}_{5}=\mathcal{O}_{6}
			=\mathcal{O}_{1,2}=\mathcal{O}_{3,4}=\mathcal{O}_{4,5}
			=\mathcal{O}_{6,1}=-2 .
		\end{aligned}
	\end{equation}
	
	\paragraph{t-deformed:}
	$I^{(t)}_{1} \,\cap\, I^{(t)}_{2}$ with $\dim(I^{(t)}_{1},I^{(t)}_{2})=(0,0)$.
	
	\begin{equation}\label{eq:Gu-t-I1}
		I_{1}^{(t)}:\quad
		\begin{aligned}
			&\mathcal{O}_{1}=\mathcal{O}_{4}=\mathcal{O}_{2,3,4}
			=2-\frac{i\sqrt{2}\,(\sqrt{t}-1)}{t^{1/4}},\\[4pt]
			&\mathcal{O}_{2}=\mathcal{O}_{3}=\mathcal{O}_{5}=\mathcal{O}_{6}
			=\mathcal{O}_{3,4}=\mathcal{O}_{6,1}
			=-2+\frac{i\sqrt{2}\,(\sqrt{t}-1)}{t^{1/4}},\\[4pt]
			&\mathcal{O}_{1,2}=\mathcal{O}_{4,5}
			=-6+\frac{2}{\sqrt{t}}
			+\frac{3i\sqrt{2}\,(\sqrt{t}-1)}{t^{1/4}}
			+2\sqrt{t},\\[4pt]
			&\mathcal{O}_{2,3}=\mathcal{O}_{5,6}=\mathcal{O}_{3,4,5}
			=4-\frac{1}{\sqrt{t}}
			-\frac{2i\sqrt{2}\,(\sqrt{t}-1)}{t^{1/4}}
			-\sqrt{t},\\[4pt]
			&\mathcal{O}_{1,2,3}
			=10+\frac{i\sqrt{2}\,(\sqrt{t}-1)^{3}}{t^{3/4}}
			-\frac{4}{\sqrt{t}}
			-\frac{3i\sqrt{2}\,(\sqrt{t}-1)}{t^{1/4}}
			-4\sqrt{t}.
		\end{aligned}
	\end{equation}
	
	\begin{equation}\label{eq:Gu-t-I2}
		I_{2}^{(t)}:\quad
		\begin{aligned}
			&\mathcal{O}_{1}=\mathcal{O}_{2}=\mathcal{O}_{3}=\mathcal{O}_{4}
			=\mathcal{O}_{5}=\mathcal{O}_{6}
			=\mathcal{O}_{3,4}=\mathcal{O}_{6,1}
			=\mathcal{O}_{2,3,4}
			=2-\frac{i\sqrt{2}\,(\sqrt{t}-1)}{t^{1/4}},\\[4pt]
			&\mathcal{O}_{1,2}=\mathcal{O}_{4,5}
			=6-\frac{2}{\sqrt{t}}
			-\frac{3i\sqrt{2}\,(\sqrt{t}-1)}{t^{1/4}}
			-2\sqrt{t},\\[4pt]
			&\mathcal{O}_{2,3}=\mathcal{O}_{5,6}=\mathcal{O}_{3,4,5}
			=4-\frac{1}{\sqrt{t}}
			-\frac{2i\sqrt{2}\,(\sqrt{t}-1)}{t^{1/4}}
			-\sqrt{t},\\[4pt]
			&\mathcal{O}_{1,2,3}
			=10+\frac{i\sqrt{2}\,(\sqrt{t}-1)^{3}}{t^{3/4}}
			-\frac{4}{\sqrt{t}}
			-\frac{3i\sqrt{2}\,(\sqrt{t}-1)}{t^{1/4}}
			-4\sqrt{t}.
		\end{aligned}
	\end{equation}

	\subsection{\texorpdfstring{$\bm{G_{x}}$ subgroup}{G_{x} subgroup}}
	
	\(G_{x}\cong \mathrm{SL}_{2}(3)\), \(|G_{x}|=24\), data \(\{0;\{3,2\},\{4,1\}\}\).
	
	\paragraph{$\bm{t=1}$:}
	$I_{1}$ with $\dim(I_{1})=0$.
	
	\begin{equation}\label{eq:Gx-I1}
		I_{1}
		=
		\Bigl\langle\,
		\mathcal{O}-2,\ \forall\,\mathcal{O}
		\,\Bigr\rangle .
	\end{equation}
	
	\paragraph{t-deformed:}
	$I^{(t)}_{1}$ with $\dim(I^{(t)}_{1})=0$.
	
	\begin{equation}\label{eq:Gx-t-I2}
		I_{2}^{(t)}:\quad
		\begin{aligned}
			&\mathcal{O}_{1}=\mathcal{O}_{2}=\mathcal{O}_{3}=\mathcal{O}_{4}
			=\mathcal{O}_{5}=\mathcal{O}_{4,5}=\mathcal{O}_{6,1}=\mathcal{O}_{3,4,5}
			=2-\frac{i\sqrt{2}\,(\sqrt{t}-1)}{t^{1/4}},\\[4pt]
			&\mathcal{O}_{6}=\mathcal{O}_{2,3}=\mathcal{O}_{3,4}
			=4-\frac{1}{\sqrt{t}}
			-\frac{2i\sqrt{2}\,(\sqrt{t}-1)}{t^{1/4}}
			-\sqrt{t},\\[4pt]
			&\mathcal{O}_{1,2}=\mathcal{O}_{2,3,4}
			=6-\frac{2}{\sqrt{t}}
			-\frac{3i\sqrt{2}\,(\sqrt{t}-1)}{t^{1/4}}
			-2\sqrt{t},\\[4pt]
			&\mathcal{O}_{5,6}
			=14+\frac{i\sqrt{2}\,(\sqrt{t}-1)^{3}}{t^{3/4}}
			-\frac{6}{\sqrt{t}}
			-\frac{5i\sqrt{2}\,(\sqrt{t}-1)}{t^{1/4}}
			-6\sqrt{t},\\[4pt]
			&\mathcal{O}_{1,2,3}
			=10+\frac{i\sqrt{2}\,(\sqrt{t}-1)^{3}}{t^{3/4}}
			-\frac{4}{\sqrt{t}}
			-\frac{3i\sqrt{2}\,(\sqrt{t}-1)}{t^{1/4}}
			-4\sqrt{t}.
		\end{aligned}
	\end{equation}

    \begin{equation}\label{eq:Gm-t-I3}
    I_{3}^{(t)}:\quad
    \begin{aligned}
    &\mathcal{O}_{1}=\mathcal{O}_{2}=\mathcal{O}_{3}=\mathcal{O}_{4}
    =\mathcal{O}_{4,5}=\mathcal{O}_{6,1}
    =-\mathcal{O}_{5}=-\mathcal{O}_{3,4,5}
    =-2-\frac{i\sqrt{2}\,(\sqrt{t}-1)}{t^{1/4}},\\[4pt]
    &\mathcal{O}_{6}=\mathcal{O}_{2,3}=\mathcal{O}_{3,4}
    =4-\frac{1}{\sqrt{t}}
    +\frac{2i\sqrt{2}\,(\sqrt{t}-1)}{t^{1/4}}
    -\sqrt{t},\\[4pt]
    &\mathcal{O}_{1,2}
    =6-\frac{2}{\sqrt{t}}
    +\frac{3i\sqrt{2}\,(\sqrt{t}-1)}{t^{1/4}}
    -2\sqrt{t},\\[4pt]
    &\mathcal{O}_{5,6}
    =14-\frac{i\sqrt{2}\,(\sqrt{t}-1)^{3}}{t^{3/4}}
    -\frac{6}{\sqrt{t}}
    +\frac{5i\sqrt{2}\,(\sqrt{t}-1)}{t^{1/4}}
    -6\sqrt{t},\\[4pt]
    &\mathcal{O}_{1,2,3}
    =-10+\frac{i\sqrt{2}\,(\sqrt{t}-1)^{3}}{t^{3/4}}
    +\frac{4}{\sqrt{t}}
    -\frac{3i\sqrt{2}\,(\sqrt{t}-1)}{t^{1/4}}
    +4\sqrt{t},\\[4pt]
    &\mathcal{O}_{2,3,4}
    =-6+\frac{2}{\sqrt{t}}
    -\frac{3i\sqrt{2}\,(\sqrt{t}-1)}{t^{1/4}}
    +2\sqrt{t}.
    \end{aligned}
    \end{equation}

    \begin{equation}\label{eq:Gm-t-I4}
    I_{4}^{(t)}:\quad
    \begin{aligned}
    &-\mathcal{O}_{1}
    =\mathcal{O}_{2}=\mathcal{O}_{3}=\mathcal{O}_{4}
    =-\mathcal{O}_{5}=-\mathcal{O}_{4,5}=-\mathcal{O}_{6,1}=-\mathcal{O}_{3,4,5}
    =2+\frac{i\sqrt{2}\,(\sqrt{t}-1)}{t^{1/4}},\\[4pt]
    &\mathcal{O}_{6}=\mathcal{O}_{2,3}=\mathcal{O}_{3,4}
    =4-\frac{1}{\sqrt{t}}
    +\frac{2i\sqrt{2}\,(\sqrt{t}-1)}{t^{1/4}}
    -\sqrt{t},\\[4pt]
    &-\mathcal{O}_{1,2}=\mathcal{O}_{2,3,4}
    =6-\frac{2}{\sqrt{t}}
    +\frac{3i\sqrt{2}\,(\sqrt{t}-1)}{t^{1/4}}
    -2\sqrt{t},\\[4pt]
    &\mathcal{O}_{5,6}
    =-14+\frac{i\sqrt{2}\,(\sqrt{t}-1)^{3}}{t^{3/4}}
    +\frac{6}{\sqrt{t}}
    -\frac{5i\sqrt{2}\,(\sqrt{t}-1)}{t^{1/4}}
    +6\sqrt{t},\\[4pt]
    &\mathcal{O}_{1,2,3}
    =-10+\frac{i\sqrt{2}\,(\sqrt{t}-1)^{3}}{t^{3/4}}
    +\frac{4}{\sqrt{t}}
    -\frac{3i\sqrt{2}\,(\sqrt{t}-1)}{t^{1/4}}
    +4\sqrt{t}.
    \end{aligned}
    \end{equation}

	\section{\texorpdfstring{$\bm{\mathcal{N}=2}$ SCFT interpretation and finite-group quotients}{\mathcal{N}=2 SCFT interpretation and finite-group quotients}}

	\vspace{0.50cm}
    In fact, interesting physical interpretation of the fixed-point subvarieties computed above stems from the nonabelian Hodge theory, where the $SL(2,\mathbb C)$ character variety of $\Sigma_2$ corresponds to the Betti moduli space of Hitchin system. The last is known to be the natural Seiberg--Witten integrable system associated with the compactification of the 6d $(2,0)$ theory on the 2d Riemann surface. For a finite subgroup $\Gamma\subset \mathrm{Mod}(\Sigma_2)$ that is realized by holomorphic automorphisms of $\Sigma_2$ \cite{Broughton'1991}, the associated quotient maps are Galois covers \cite{CecottiDelZotto:2015GaloisCovers} with specific ramification data (indicated in our figures). The $\Gamma$ fixed points of the (classical or $t$--deformed) character variety correspond to equivariant flat connections.

    Majority of the derived surfaces appear to be genus zero, for which the fixed locus can be identified with a relative character variety of a punctured sphere (irregular character variety). The studied cases are consistent with the structure of rank-one and rank-two Hitchin integrable systems with multiple punctures (\textit{e.g.} $n =4, 5$). This makes the $2$-- and $4$--dimensional ideals to be natural candidates for the Coulomb branch geometries of 4d $\mathcal N=2$ theories. In the given context, the compactification of the six-dimensional $\mathcal N=(2,0)$ theory of type $\mathfrak{g}$ on a punctured Riemann surface with suitable irregular singularities leads to the class-$\mathcal{S}$ construction \cite{Gaiotto2009,GMN2010,Xie2012}, which corresponds to the class of Argyres-Douglas (AD) Supersymmetric CFTs \cite{ArgyresDouglas1995, Cecotti:2020Swampland}  (wild Hitchin type). Specifically, it is of particular interest to consider the 2d, 4d singular subvarieties ($ G_{b},\; G_{f},\; G_{e},\; G_{k},\; G_{n},\; G_{s}$) or their reductions and investigate the emerging SCFTs, their characteristic spectral curves (SW geometry) and other properties. It is also important to investigate the subvariety equivalence phenomena at the associated AD SCFT level, i.e. when the genus and number of irregularities transforms for equivalent subvarieties, whose subgroups are different $ G_{b}(G_{f}),\; G_{h}(G_{o}),\; G_{i}(G_{p}),\; G_{k_{2}}(G_{s}),\; G_{r}(G_{w}) $, which is a subject of the upcoming work. 

    In addition, from the F-theory side, the emerging meromorphic Hitchin systems provide the local description of 7-brane gauge sectors. A related problem suggested by our results is to compare each finite-group quotient with the corresponding symmetry reduced Hitchin system on the quotient curve. One may further ask whether the corresponding $ \{q,t\} $-deformation leads to novel quantum reduced varieties and whether it provides new (or reproduces) protected observables of the associated 4d theory, such as wall-crossing data and defect operator algebras.

    \paragraph{Acknowledgments.} We would like to thank Alexander Belavin, Sergio Cecotti, Dmitry Orlov, Chris Brav, Mykola Dedushenko and Yankun Ma for stimulating discussions and useful comments. The work of S.A. was supported by the Beijing Natural Science Foundation grant IS25025. Research of AP was supported by Beijing Natural Science Foundation grant IS25017 and Russian Science Foundation grant \href{https://ias.rscf.ru/?/user/doc/a.w.p.2025.111.main/20162011}{RSCF-25-72-10177}. Part of this work was completed during a research visit at the International Laboratory for Mirror Symmetry and Automorphic Forms, NRU HSE and AP express gratitude to Valery Gritsenko and Mikhail Alfimov for hospitality.

\newpage
	\appendix
	
	\section{Genus two DAHA and Twist actions}

	\vspace{0.50cm}
	\subsection{Generators and $J$-relations of $\cl{A}_{1,t}$} 
	For later use in the fixed-point analysis, we record the mapping class group action on the
	genus-two DAHA generators and its specialization to the classical \(t=1\) fiber. The genus two DAHA can be spanned by the following 15 generators 
	\begin{equation} 
		\mathcal O_{1},\dots,\mathcal O_{6},\qquad
		\mathcal O_{1,2},\mathcal O_{2,3},\mathcal O_{3,4},\mathcal O_{4,5},\mathcal O_{5,6},\mathcal O_{6,1},
		\qquad
		\mathcal O_{1,2,3},\mathcal O_{2,3,4},\mathcal O_{3,4,5}.
	\end{equation} 
	
	The commutative classical limit is denoted by \(\mathcal A_{q=1,t}\), and its further specialization at
	\(t=1\) is 
	\begin{equation}
		\mathcal A_{q=t=1}
		\cong
		\mathcal O\!\left(Hom(\pi_{1}(\Sigma_{2}),SL(2,\mathbb C))\right)^{SL(2,\mathbb C)}.
	\end{equation}
	
	\nd The $\cl{O}_{i}$ and $O_{i,i+1}$ are orbit 6 and $\cl{O}_{i,i+1,i+2}$ is orbit 3 generators 
	\begin{equation}
		\mathcal O_{4,5,6}=\mathcal O_{1,2,3},\qquad
		\mathcal O_{5,6,1}=\mathcal O_{2,3,4},\qquad
		\mathcal O_{6,1,2}=\mathcal O_{3,4,5}.
	\end{equation} 
    The complete $ J_{q,t} $-relations are given by 
	\begin{equation}
		\begin{aligned}
			& q^{-1/2}\,O_{i+2}O_{i+4}
			+q^{1/2}\,O_{i+3}O_{i+2,i+3,i+4} \\
			& -O_{i+2,i+3}O_{i+3,i+4}-\Bigl(q^{1/2}t^{-1/2}+q^{-1/2}t^{1/2}\Bigr)O_i = 0, \\[2ex] 
			& q^{-1/2}\,O_{i+3,i+4}O_{i+1,i+2,i+3} -q^{-1}\,O_{i+3}O_{i+5,i}
			- O_{i+4}O_{i+1,i+2} \\
			& -\Bigl(q^{1/2}t^{-1/2}+q^{-1/2}t^{1/2}\Bigr)
			\Bigl(O_{i,i+1}-q^{-1/4}O_iO_{i+1}\Bigr) = 0, \\[2ex]
			& O_{i,i+1,i+2}O_{i+1,i+2,i+3} -q^{1/2}\,O_{i+1,i+2}O_{i+4,i+5}
			- q^{-1/2}\,O_iO_{i+3} \\ 
			& -\Bigl(q^{1/2}t^{-1/2}+q^{-1/2}t^{1/2}\Bigr)\!
			\Bigl(
			-(q-1+q^{-1})\,O_{i+2,i+3,i+4} \\ 
			&+ q^{-3/4}O_{i+1}O_{i+5,i}
			+ q^{3/4}O_{i+5}O_{i,i+1}
			- O_iO_{i+1}O_{i+5}
			\Bigr) = 0 \,.\\[1ex]
			&\mathcal{O}_{1,2,3}\mathcal{O}_{2,3,4}\mathcal{O}_{3,4,5}
			-\bigl(
			\mathcal{O}_{1}\mathcal{O}_{4}\mathcal{O}_{3,4,5}
			+\mathcal{O}_{2}\mathcal{O}_{5}\mathcal{O}_{1,2,3}
			+\mathcal{O}_{3}\mathcal{O}_{6}\mathcal{O}_{2,3,4}
			\bigr) \\
			&\quad
			+\frac12\bigl(t^{1/2}+t^{-1/2}\bigr)\bigl(
			\mathcal{O}_{1}\mathcal{O}_{6}\mathcal{O}_{6,1}
			+\mathcal{O}_{2}\mathcal{O}_{1}\mathcal{O}_{1,2}
			+\mathcal{O}_{3}\mathcal{O}_{2}\mathcal{O}_{2,3} \\
			&\quad +\mathcal{O}_{4}\mathcal{O}_{3}\mathcal{O}_{3,4}
			+\mathcal{O}_{5}\mathcal{O}_{4}\mathcal{O}_{4,5}
			+\mathcal{O}_{6}\mathcal{O}_{5}\mathcal{O}_{5,6}
			\bigr) - \bigl(
			\mathcal{O}_{1}\mathcal{O}_{3}\mathcal{O}_{5}
			+\mathcal{O}_{2}\mathcal{O}_{4}\mathcal{O}_{6}
			\bigr) \\
			&\quad
			-\frac12\bigl(t^{1/2}+t^{-1/2}\bigr)\bigl(
			\mathcal{O}_{1,2}^{2}
			+\mathcal{O}_{2,3}^{2}
			+\mathcal{O}_{3,4}^{2}
			+\mathcal{O}_{4,5}^{2}
			+\mathcal{O}_{5,6}^{2}
			+\mathcal{O}_{6,1}^{2}
			\bigr)
			+\bigl(t^{1/2}+t^{-1/2}\bigr)^3=0.
		\end{aligned}
	\end{equation}
	In the commutative limit the normal-ordering relations are $ \mathcal O_{I}\mathcal O_{J}=\mathcal O_{J}\mathcal O_{I} $ and $ t=1 $ fiber leads to the specialized $J$-relations and one specialized Casimir relation.

	\subsection{Mapping Class Group $ Mod(\Sigma_{2}) $ and automorphisms}The mapping class group action is generated by the Dehn twist \(d_{1}\) and the order-six automorphism
	\begin{equation}
		I=
		(\mathcal O_{1},\mathcal O_{2},\mathcal O_{3},\mathcal O_{4},\mathcal O_{5},\mathcal O_{6})
		(\mathcal O_{1,2},\mathcal O_{2,3},\mathcal O_{3,4},\mathcal O_{4,5},\mathcal O_{5,6},\mathcal O_{6,1})
		(\mathcal O_{1,2,3},\mathcal O_{2,3,4},\mathcal O_{3,4,5}),
	\end{equation}
	together with 
	\begin{equation}
		d_{i}=I^{\,i-1}d_{1}I^{\,1-i},\qquad i=1,\dots,5.
	\end{equation}
	The braid and commutation relations are
	\begin{equation}
		d_{i}d_{i+1}d_{i}=d_{i+1}d_{i}d_{i+1}\qquad (1\le i\le 4),
		\qquad
		d_{i}d_{j}=d_{j}d_{i}\qquad (|i-j|>1),
	\end{equation}
	and \(I^{6}=1\). In fact, the hyperelliptic involution acts trivially in this representation, so
	\begin{equation}
		H:=d_{1}d_{2}d_{3}d_{4}d_{5}d_{5}d_{4}d_{3}d_{2}d_{1}=1 \,, 
	\end{equation}
	which in the present framework also corresponds to the composite twist $\zeta_{0}$. 
	
	\subsection*{Quantum and classical \(d_{1}\)-actions on \(\mathcal A_{q,t}\)}
	
	In the quantum algebra \(\mathcal A_{q,t}\), the Dehn twist \(d_{1}\) and its inverse act by
	\begin{equation}
		\begin{cases}
			\mathcal O_{2}\mapsto q^{1/4}\mathcal O_{1}\mathcal O_{2}-q^{1/2}\mathcal O_{1,2},\\
			\mathcal O_{6}\mapsto \mathcal O_{6,1},\\
			\mathcal O_{1,2}\mapsto \mathcal O_{2},\\
			\mathcal O_{2,3}\mapsto q^{1/4}\mathcal O_{1}\mathcal O_{2,3}-q^{1/2}\mathcal O_{1,2,3},\\
			\mathcal O_{5,6}\mapsto \mathcal O_{2,3,4},\\
			\mathcal O_{6,1}\mapsto q^{1/4}\mathcal O_{1}\mathcal O_{6,1}-q^{1/2}\mathcal O_{6},\\
			\mathcal O_{1,2,3}\mapsto \mathcal O_{2,3},\\
			\mathcal O_{2,3,4}\mapsto q^{1/4}\mathcal O_{1}\mathcal O_{2,3,4}-q^{1/2}\mathcal O_{5,6},
		\end{cases} 
		\qquad 
		d_{1}^{-1}:
		\begin{cases}
			\mathcal O_{2}\mapsto \mathcal O_{1,2},\\
			\mathcal O_{6}\mapsto q^{-1/4}\mathcal O_{1}\mathcal O_{6}-q^{-1/2}\mathcal O_{6,1},\\
			\mathcal O_{1,2}\mapsto q^{-1/4}\mathcal O_{1}\mathcal O_{1,2}-q^{-1/2}\mathcal O_{2},\\
			\mathcal O_{2,3}\mapsto \mathcal O_{1,2,3},\\
			\mathcal O_{5,6}\mapsto q^{-1/4}\mathcal O_{1}\mathcal O_{5,6}-q^{-1/2}\mathcal O_{2,3,4},\\
			\mathcal O_{6,1}\mapsto \mathcal O_{6},\\
			\mathcal O_{1,2,3}\mapsto q^{-1/4}\mathcal O_{1}\mathcal O_{1,2,3}-q^{-1/2}\mathcal O_{2,3},\\
			\mathcal O_{2,3,4}\mapsto \mathcal O_{5,6},
		\end{cases}
	\end{equation}
	while the action on remaining generators is trivial. In the classical limit \(q=1\) the analogous formulas define an automorphism of the commutative algebra
	\(\mathcal A_{q=1,t}\):
	\begin{equation}
		d_{1}:
		\begin{cases}
			\mathcal O_{2}\mapsto \mathcal O_{1}\mathcal O_{2}-\mathcal O_{1,2},\\
			\mathcal O_{6}\mapsto \mathcal O_{6,1},\\
			\mathcal O_{1,2}\mapsto \mathcal O_{2},\\
			\mathcal O_{2,3}\mapsto \mathcal O_{1}\mathcal O_{2,3}-\mathcal O_{1,2,3},\\
			\mathcal O_{5,6}\mapsto \mathcal O_{2,3,4},\\
			\mathcal O_{6,1}\mapsto \mathcal O_{1}\mathcal O_{6,1}-\mathcal O_{6},\\
			\mathcal O_{1,2,3}\mapsto \mathcal O_{2,3},\\
			\mathcal O_{2,3,4}\mapsto \mathcal O_{1}\mathcal O_{2,3,4}-\mathcal O_{5,6},
		\end{cases}
		\qquad 
		d_{1}^{-1}:
		\begin{cases}
			\mathcal O_{2}\mapsto \mathcal O_{1,2},\\
			\mathcal O_{6}\mapsto \mathcal O_{1}\mathcal O_{6}-\mathcal O_{6,1},\\
			\mathcal O_{1,2}\mapsto \mathcal O_{1}\mathcal O_{1,2}-\mathcal O_{2},\\
			\mathcal O_{2,3}\mapsto \mathcal O_{1,2,3},\\
			\mathcal O_{5,6}\mapsto \mathcal O_{1}\mathcal O_{5,6}-\mathcal O_{2,3,4},\\
			\mathcal O_{6,1}\mapsto \mathcal O_{6},\\
			\mathcal O_{1,2,3}\mapsto \mathcal O_{1}\mathcal O_{1,2,3}-\mathcal O_{2,3},\\
			\mathcal O_{2,3,4}\mapsto \mathcal O_{5,6}.
		\end{cases}
	\end{equation}
	The same commutative formulas act on the fiber \(\mathcal A_{q=t=1}\). 
	
	\subsection{Finite Subgroup $\zeta$-twist actions}
	
	As it was indicated in \cite{Broughton'1991, NakamuraNakanishi'2018}, the $ Mod(\Sigma_{2}) $ subgroup actions are defined through $\zeta$ composite twists, which action on the genus two DAHA generators $\cl{O}$ derives: \\
	
	The $\zeta_{0}$ defines hyperinvolution, hence as indicated above, acts trivially on the DAHA generators $\zeta_{0}(\cl{O}) = \cl{O}$ . \\
	
	\begin{equation}\label{eq:zeta1-action}
		\zeta_{1}:\quad
		\begin{array}{rcl@{\qquad}rcl@{\qquad}rcl}
			\mathcal{O}_{1} &\to& \mathcal{O}_{2} &
			\mathcal{O}_{1,2} &\to& \mathcal{O}_{2,3} &
			\mathcal{O}_{1,2,3} &\to& \mathcal{O}_{2,3,4} \\
			
			\mathcal{O}_{2} &\to& \mathcal{O}_{3} &
			\mathcal{O}_{2,3} &\to& \mathcal{O}_{3,4} &
			\mathcal{O}_{2,3,4} &\to& \mathcal{O}_{3,4,5} \\
			
			\mathcal{O}_{3} &\to& \mathcal{O}_{4} &
			\mathcal{O}_{3,4} &\to& \mathcal{O}_{4,5} &
			\mathcal{O}_{3,4,5} &\to& \mathcal{O}_{1,2,3} \\
			
			\mathcal{O}_{4} &\to& \mathcal{O}_{5} &
			\mathcal{O}_{4,5} &\to& \mathcal{O}_{5,6} & & \\
			
			\mathcal{O}_{5} &\to& \mathcal{O}_{6} &
			\mathcal{O}_{5,6} &\to& \mathcal{O}_{6,1} & & \\
			
			\mathcal{O}_{6} &\to& \mathcal{O}_{1} &
			\mathcal{O}_{6,1} &\to& \mathcal{O}_{1,2} & &
		\end{array}
	\end{equation}

	\begin{equation}\label{eq:zeta2-action}
		\zeta_{2}:\quad
		\begin{array}{c}
			\begin{array}{rcl}
				\mathcal{O}_{1} &\to& \mathcal{O}_{2}\\
				\mathcal{O}_{2} &\to& \mathcal{O}_{1}\mathcal{O}_{2}-\mathcal{O}_{1,2}\\
				\mathcal{O}_{3} &\to& \mathcal{O}_{5}\mathcal{O}_{6}-\mathcal{O}_{5,6}\\
				\mathcal{O}_{4} &\to& \mathcal{O}_{5}\\
				\mathcal{O}_{5} &\to& \mathcal{O}_{4,5}\\
				\mathcal{O}_{6} &\to& \mathcal{O}_{2,3}
			\end{array}
			\qquad
			\begin{array}{rcl}
				\mathcal{O}_{1,2} &\to& \mathcal{O}_{1}(\mathcal{O}_{2}^{2}-1)-\mathcal{O}_{2}\mathcal{O}_{1,2}\\
				\mathcal{O}_{2,3} &\to& \mathcal{O}_{3}\mathcal{O}_{4}-\mathcal{O}_{3,4}\\
				\mathcal{O}_{3,4} &\to& (\mathcal{O}_{5}^{2}-1)\mathcal{O}_{6}-\mathcal{O}_{5}\mathcal{O}_{5,6}\\
				\mathcal{O}_{4,5} &\to& \mathcal{O}_{4}\\
				\mathcal{O}_{5,6} &\to& \mathcal{O}_{2,3}\mathcal{O}_{4,5}-\mathcal{O}_{6,1}\\
				\mathcal{O}_{6,1} &\to& \mathcal{O}_{3}
			\end{array}
			\\[6pt]
			\begin{array}{rcl}
				\mathcal{O}_{1,2,3} &\to& \mathcal{O}_{4}\mathcal{O}_{2,3}-\mathcal{O}_{2,3,4}\\
				\mathcal{O}_{2,3,4} &\to& \mathcal{O}_{3}\mathcal{O}_{4,5}-\mathcal{O}_{3,4,5}\\
				\mathcal{O}_{3,4,5} &\to&
				\tfrac12\!\left(\mathcal{O}_{2}\mathcal{O}_{5}
				+\mathcal{O}_{3,4}\mathcal{O}_{6,1}
				-\mathcal{O}_{2,3,4}\mathcal{O}_{3,4,5}\right)
			\end{array}
		\end{array}
	\end{equation}

	\begin{equation}\label{eq:zeta3-action}
		\zeta_{3}:\quad
		\begin{array}{c}
			\begin{array}{rcl}
				\mathcal{O}_{1} &\to& \mathcal{O}_{1}\mathcal{O}_{2}-\mathcal{O}_{1,2}\\
				\mathcal{O}_{2} &\to& \mathcal{O}_{3}\\
				\mathcal{O}_{3} &\to& \mathcal{O}_{4}\\
				\mathcal{O}_{4} &\to& \mathcal{O}_{5}\mathcal{O}_{6,1}-\mathcal{O}_{2,3,4}\\
				\mathcal{O}_{5} &\to& \mathcal{O}_{6,1}\\
				\mathcal{O}_{6} &\to& -\mathcal{O}_{6}+\mathcal{O}_{1}\mathcal{O}_{6,1}
			\end{array}
			\qquad
			\begin{array}{rcl}
				\mathcal{O}_{1,2} &\to& \mathcal{O}_{1}\mathcal{O}_{2,3}-\mathcal{O}_{1,2,3}\\
				\mathcal{O}_{2,3} &\to& \mathcal{O}_{3,4}\\
				\mathcal{O}_{3,4} &\to& -\mathcal{O}_{2,3}+\mathcal{O}_{4,5}\mathcal{O}_{6,1}\\
				\mathcal{O}_{4,5} &\to& \mathcal{O}_{5}\\
				\mathcal{O}_{5,6} &\to& -\mathcal{O}_{6}\mathcal{O}_{6,1}
				+\mathcal{O}_{1}\bigl(-1+\mathcal{O}_{6,1}^{2}\bigr)\\
				\mathcal{O}_{6,1} &\to& \mathcal{O}_{3,4,5}
			\end{array}
			\\[6pt]
			\begin{array}{rcl}
				\mathcal{O}_{1,2,3} &\to& -\mathcal{O}_{5,6}+\mathcal{O}_{1}\mathcal{O}_{2,3,4}\\
				\mathcal{O}_{2,3,4} &\to& -\mathcal{O}_{2}+\mathcal{O}_{6,1}\mathcal{O}_{3,4,5}\\
				\mathcal{O}_{3,4,5} &\to& \mathcal{O}_{4,5}
			\end{array}
		\end{array}
	\end{equation}

	\begin{equation}\label{eq:zeta4-action}
		\zeta_{4}:\quad
		\begin{array}{c}
			\begin{array}{rcl}
				\mathcal{O}_{1} &\to& \mathcal{O}_{2}\\
				\mathcal{O}_{2} &\to& \mathcal{O}_{3}\\
				\mathcal{O}_{3} &\to& \mathcal{O}_{4}\\
				\mathcal{O}_{4} &\to& \mathcal{O}_{5}\mathcal{O}_{6}-\mathcal{O}_{5,6}\\
				\mathcal{O}_{5} &\to& \mathcal{O}_{6}\\
				\mathcal{O}_{6} &\to& \mathcal{O}_{6,1}
			\end{array}
			\qquad
			\begin{array}{rcl}
				\mathcal{O}_{1,2} &\to& \mathcal{O}_{2,3}\\
				\mathcal{O}_{2,3} &\to& \mathcal{O}_{3,4}\\
				\mathcal{O}_{3,4} &\to& \mathcal{O}_{6}\mathcal{O}_{4,5}-\mathcal{O}_{1,2,3}\\
				\mathcal{O}_{4,5} &\to& \mathcal{O}_{5}\\
				\mathcal{O}_{5,6} &\to& -\mathcal{O}_{1}+\mathcal{O}_{6}\mathcal{O}_{6,1}\\
				\mathcal{O}_{6,1} &\to& \mathcal{O}_{3,4,5}
			\end{array}
			\\[6pt]
			\begin{array}{rcl}
				\mathcal{O}_{1,2,3} &\to& \mathcal{O}_{2,3,4}\\
				\mathcal{O}_{2,3,4} &\to& -\mathcal{O}_{1,2}+\mathcal{O}_{6}\mathcal{O}_{3,4,5}\\
				\mathcal{O}_{3,4,5} &\to& \mathcal{O}_{4,5}
			\end{array}
		\end{array}
	\end{equation}

\newpage
\bigskip
\noindent\rule{\textwidth}{0.4pt}

\section*{Affiliations}

\noindent
\textbf{Semeon Arthamonov}\\
Beijing Institute of Mathematical Sciences and Applications (BIMSA)\\
Huairou District, Beijing 101408, China\\
\textit{E-mail:} \href{mailto:arthamonov@bimsa.cn}{arthamonov@bimsa.cn}

\medskip
\medskip

\noindent
\textbf{Anton Pribytok}\\
Beijing Institute of Mathematical Sciences and Applications (BIMSA)\\
Huairou District, Beijing 101408, China\\[2pt]
Steklov Mathematical Institute of Russian Academy of Sciences\\
Gubkina str. 8, 119991 Moscow, Russia\\[2pt]
Faculty of Mathematics, National Research University Higher School of Economics, Usacheva str. 6, 119048 Moscow, Russia\\[2pt]
\textit{E-mail:} 
\href{mailto:antonspribitoks@bimsa.cn}{antonspribitoks@bimsa.cn}

\newpage
\bibliographystyle{alpha}
\bibliography{references} 	
	
\end{document}